\documentclass[12pt,fleqn]{article}
\usepackage{graphicx}
\usepackage{cite}

\newcommand{\psla}{p\!\!\!/\,}
\newcommand{\ksla}{k\!\!\!/\,}
\newcommand{\Dsla}{D\!\!\!\!/\,\,}

\newcommand{\Toneep}{T_1^\varepsilon}
\newcommand{\Stwo}{S_2}
\newcommand{\OepStwo}{S_2^\varepsilon}

\newcommand{\Dthree}{D_3}

\newcommand{\Bfour}{B_4}

\newcommand{\zetathree}{\zeta(3)}

\newcommand{\fracown}[2]{{\textstyle \frac{#1}{#2}}}

\def\theequation{\thesection.\arabic{equation}}
\newcommand{\newsection}[1]{\section{#1}\setcounter{equation}{0}}
\newcommand{\newappendix}[1]{\section*{#1}\setcounter{equation}{0}}

\def\be{\begin{equation}}
\def\ee{\end{equation}}
\def\bea{\begin{eqnarray}}
\def\eea{\end{eqnarray}}
\def\nnb{\nonumber}
\def\bbuildrel#1_#2^#3{\mathrel{\mathop{\kern 0pt#1}\limits_{#2}^{#3}}}
\def\slash#1{\setbox0=\hbox{$#1$}#1\hskip-\wd0\dimen0=5pt\advance
       \dimen0 by-\ht0\advance\dimen0 by\dp0\lower0.5\dimen0\hbox
         to\wd0{\hss\sl/\/\hss}}

\newcommand{\scs}{\scriptscriptstyle}
\newcommand{\f}{\frac}
\newcommand{\fm}[2]{{\textstyle \frac{#1}{#2}}}

\newcommand{\al}{\widetilde{\alpha}_{\mathrm s}}
\newcommand{\e}{\epsilon}
\newcommand{\lw}{\ln  \f{\mu_0^2}{M_W^2}} 
\newcommand{\lwsq}{\ln^2\f{\mu_0^2}{M_W^2}}

\begin{document}

\begin{titlepage}

\begin{flushright}
DESY 04-001\\
IFT-01/2004\\
ZU-TH-1/04\\

hep-ph/0401041\\[2cm]
\end{flushright}

\begin{center}
\setlength {\baselineskip}{0.3in} 
{\bf\Large 
  Three-loop matching of the dipole operators\\ for $b\to s\gamma$ and $b\to s g$ }\\[2cm]
\setlength {\baselineskip}{0.2in}
{\large  Miko{\l}aj Misiak$^{1,3}$ and Matthias Steinhauser$^2$}\\[5mm]
$^1$~{\it Institute of Theoretical Physics, Warsaw University,\\
         Ho\.za 69, PL-00-681 Warsaw, Poland.}\\[5mm] 
$^2$~{\it II. Institut f\"ur Theoretische Physik, Universit\"at Hamburg,\\
          Luruper Chaussee 149, D-22761, Hamburg, Germany.}\\[5mm]
$^3$~{\it Institut f\"ur Theoretische Physik, Universit\"at Z\"urich,\\
          Winterthurerstrasse 190, CH-8057, Z\"urich, Switzerland.}\\[3cm]
{\bf Abstract}\\[5mm]
\end{center} 
\setlength{\baselineskip}{0.2in} 

We evaluate the three-loop matching conditions for the dimension-five
operators that are relevant for the $b \to s \gamma$ decay. Our
calculation completes the first out of three steps (matching, mixing
and matrix elements) that are necessary for finding the
next-to-next-to-leading QCD corrections to this process. All such
corrections must be calculated in view of the ongoing accurate
measurements of the $\bar{B} \to X_s \gamma$ branching ratio.

\end{titlepage}

\newsection{Introduction \label{sec:intro}}

The inclusive weak radiative $\bar{B}$-meson decay is known to be a
sensitive probe of new physics.  Its branching ratio has been measured
by CLEO \cite{Chen:2001fj}, ALEPH \cite{Barate:1998vz}, BELLE
\cite{Abe:2001hk} and BABAR \cite{Aubert:2002pd}.  The experimental
world average\cite{Jessop.2002}
\be \label{waverage}
BR[\bar{B} \to X_s \gamma,~~( E_{\gamma} > \fm{1}{20} m_b) ] 
= ( 3.34 \pm 0.38 ) \times 10^{-4}
\ee
agrees with the Standard Model (SM) predictions
\cite{Gambino:2001ew,Buras:2002tp}
\bea \label{thSM}
BR[\bar{B} \to X_s \gamma,~~ (E_{\gamma} > 1.6~{\rm GeV}) ] 
                                            &=& ( 3.57 \pm 0.30 ) \times 10^{-4},\\
BR[\bar{B} \to X_s \gamma,~~( E_{\gamma} > \fm{1}{20} m_b) ] &\simeq& 3.70 \times 10^{-4}.
\eea
Such a good agreement implies constraints on a variety of extensions
of the SM, including the Minimal Supersymmetric Standard Model with
superpartner masses ranging up to several hundreds GeV. These
constraints are expected to be crucial for identification of possible
new physics signals at the Tevatron, LHC and other high-energy
colliders. However, any future increase of their power depends on
whether the theoretical calculations manage to follow the improving
accuracy of the experimental determinations of 
BR$[\bar{B} \to X_s \gamma]$.

\begin{figure}[h]
\begin{center}
\includegraphics[width=5cm,angle=0]{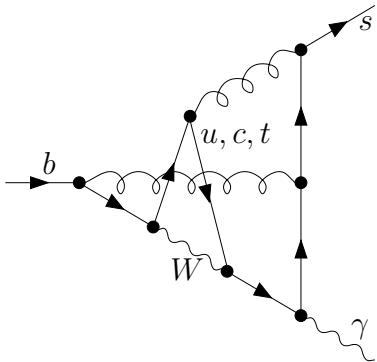}
\end{center}
\vspace*{-54mm}
\hspace*{106mm} $s$\\[1cm]
\hspace*{85mm} $u,c,t$\\[-1mm]
\hspace*{64mm} $b$\\[9mm]
\hspace*{81mm} $W$\\[2mm]
\hspace*{105mm} $\gamma$\\[-5mm]
\begin{center}
\caption{\sf One of the ${\cal O}(10^3)$ three-loop diagrams that we have calculated.}
\label{fig:sampleSM}
\end{center}
\vspace*{-6mm}
\end{figure}

As pointed out more than two years ago \cite{Gambino:2001ew}, the main
theoretical uncertainty in the SM prediction for BR$[\bar{B} \to X_s
\gamma]$ originates from the perturbative calculation of the $b \to s
\gamma$ amplitude.  It is manifest when one considers the charm-quark
mass renormalization ambiguity \cite{Gambino:2001ew} in the two-loop,
Next-to-Leading Order (NLO) QCD corrections to this amplitude
\cite{Greub:1996tg,Buras:2002tp}. The only method for removing this
ambiguity is calculating the three-loop, Next-to-Next-to-Leading Order
(NNLO) QCD corrections. A sample NNLO diagram is shown in
Fig.~\ref{fig:sampleSM}.

Since $m_b \ll M_W$, such diagrams are most conveniently calculated
using an effective field theory language. The electroweak-scale
contributions are encoded into the {\em matching conditions} for the
Wilson coefficients, while the $b$-quark-scale contributions are seen
as {\em matrix elements} of several flavour-changing operators. Large
logarithms $\ln(M_W^2/m_b^2)$ are resummed using the effective theory
Renormalization Group Equations (RGE) that result from the operator
{\em mixing} under renormalization.

The matching conditions and the matrix elements yield a
renormalization-scheme independent contribution to the amplitude only
after combining them together. Thus, both of them need to be evaluated
to the NNLO. It is impossible to remove the charm-quark mass ambiguity
by calculating the matrix elements only, even though the matching
conditions are $m_c$-independent.

In this paper, we present our calculation of the three-loop matching
conditions for the dipole operators 
$(\bar{s}_L \sigma^{\mu \nu}     b_R) F_{\mu \nu}$
and 
$(\bar{s}_L \sigma^{\mu \nu} T^a b_R) G_{\mu \nu}^a$.
All the other matching conditions that are relevant for $b \to s
\gamma$ at the NNLO originate from two-loop diagrams only, and were
calculated several years ago~\cite{Bobeth:1999mk}. Thus, our work
completes the first (matching) step of the full ${\cal O}(\alpha_s^2)$
analysis of the considered process.

The long history of the lower-order (${\cal O}(1)$ and ${\cal
O}(\alpha_s)$) analyses has been summarized in
Ref.~\cite{Buras:2002er}. As far as the NNLO calculations are
concerned, fermion-loop contributions to the three-loop matrix element
of the current-current operator $(\bar{s}c)_{\scs V-A}(\bar{c}b)_{\scs
V-A}$ are already known \cite{Bieri:2003ue}.  Three-loop anomalous
dimensions of all the four-quark operators will soon be published
\cite{GGH04}. Work at the remaining anomalous dimensions and matrix
elements is in progress.

In our present three-loop matching computation, we follow the procedure
outlined in Ref.~\cite{Bobeth:1999mk}.  All the necessary diagrams are
evaluated off-shell, after expanding them in external momenta. The
spurious infrared divergences generated by the expansion are regulated
dimensionally.  They cancel out in the matching equation, i.e. in the
difference between the full SM and the effective theory off-shell
amplitudes.

The scalar three-loop integrals are evaluated with the help of the
package {\tt MATAD} \cite{Steinhauser:2000ry} designed for calculating
vacuum diagrams.  The fact that {\tt MATAD} can deal with a single
non-vanishing mass only is not an obstacle against taking into account
the actually different masses of the $W$ boson and the top
quark. Expansions starting from $m_t=M_W$ and $m_t \gg M_W$ allow us
to accurately determine the three-loop matching conditions for the
physical values of $m_t$ and $M_W$.

Our paper is organized as follows. In Section~\ref{sec:eff.theory}, we
introduce the effective theory and give all the necessary
renormalization constants.  In Section~\ref{sec:SMbare}, the
unrenormalized one-, two- and three-loop SM amplitudes for $b \to s
\gamma$ and $b \to sg$ are presented. Section~\ref{sec:SMren} is
devoted to a discussion of the SM counterterms. The matching procedure
is described in Section \ref{sec:matching}. Explicit expressions for
the resulting Wilson coefficients are given in
Section~\ref{sec:results}. We conclude in
Section~\ref{sec:conclusions}.  Appendix~A contains exact expressions for
the coefficients of the expansions in $(1-M_W^2/m_t^2)$ and
$M_W^2/m_t^2$.

\newsection{The effective theory \label{sec:eff.theory}}

Since our approach follows Ref.~\cite{Bobeth:1999mk} very closely, we
shall not repeat all the details given there. While the present
article is self-contained as far as the notation is concerned,
Sections 2 and 5 of Ref.~\cite{Bobeth:1999mk} are referred to for
pedagogical explanations.

The effective theory that we consider arises from the SM after
decoupling of the heavy electroweak bosons and the top quark. Its
off-shell Lagrangian reads
\be \label{Leff}
{\cal L}_{\rm eff} = {\cal L}_{\scs {\rm QCD} \times {\rm QED}}(u,d,s,c,b) 
+ \f{4 G_F}{\sqrt{2}} \sum_{i,j}  
\left[ V^*_{cs} V_{cb} C^c_i \; + \; V^*_{ts} V_{tb} C^t_i \right] Z_{ij} P_j,
\ee
where $G_F$ is the Fermi constant and $V$ stands for the
Cabibbo-Kobayashi-Maskawa (CKM) matrix. The operators $P_j$ can
be found in Eqs.~(2), (73) and (101) of
Ref.~\cite{Bobeth:1999mk}.\footnote{
For simplicity, we set $V_{ub}$ to zero here, which makes irrelevant
the operators $P_j^u$ from Ref.~\cite{Bobeth:1999mk}. The operators
$P_j^c$ from that paper are denoted by $P_j$ here. Our final results
are insensitive to whether $V_{ub}$ vanishes or not.}
The ones that are relevant for our present matching computation read
\bea 
P_1 &=& (\bar{s}_L \gamma_{\mu} T^a c_L) (\bar{c}_L \gamma^{\mu} T^a b_L),\nnb\\
P_2 &=& (\bar{s}_L \gamma_{\mu}     c_L) (\bar{c}_L \gamma^{\mu}     b_L),\nnb\\
P_4 &=& (\bar{s}_L \gamma_{\mu} T^a b_L) \sum_{q=u,d,s,c,b} 
        (\bar{q}\gamma^{\mu} T^a q),\nnb\\
P_7 &=&  Z_g^{-2}\; \f{e m_b}{g^2} (\bar{s}_L \sigma^{\mu \nu}     b_R) F_{\mu \nu},  \nnb\\
P_8 &=&  Z_g^{-2}\; \f{m_b}{g} (\bar{s}_L \sigma^{\mu \nu} T^a b_R) G_{\mu \nu}^a,\nnb\\
P_{11} &=& (\bar{s}_L \gamma_{\mu}\gamma_{\nu}\gamma_{\rho} T^a c_L)
           (\bar{c}_L \gamma^{\mu}\gamma^{\nu}\gamma^{\rho} T^a b_L) -16 P_1. 
\label{oper}
\eea
Their Wilson coefficients can be perturbatively expanded as follows
\begin{eqnarray}
  C^Q_i &=& C^{Q(0)}_i 
  + \al C^{Q(1)}_i  
  + \al^2 C^{Q(2)}_i  
  + \al^3 C^{Q(3)}_i  
  + {\cal O}(\al^4), \hspace{1cm} Q = c,t
  \,.
  \label{eq:expanded.coeffs}
\end{eqnarray}
where $\al = \alpha_s/(4\pi) = g^2/(4\pi)^2$ and $C^{Q(n)}$ originate
from $n$-loop matching conditions. We neglect the ${\cal
O}(\alpha_{\rm em})$ corrections to the r.h.s. of the above equation
as well as additional operators that arise at higher orders in the
electroweak interactions.

The goal of the present paper is finding $C^{Q(3)}_7$ and $C^{Q(3)}_8$
at the renormalization scale \linebreak $\mu_0 \sim (m_t {\rm~or~}
M_W)$. As we shall see, it is convenient to consider different scales
$\mu_0$ for $Q=c$ and $Q=t$. This is the reason why we refrain from
applying unitarity of the CKM matrix throughout the paper.

The renormalization constants $Z_{ij}$ that enter Eq.~(\ref{Leff}) are
all known to sufficiently high orders from previous calculations
\cite{Chetyrkin:1996vx,Gambino:2003zm}. The ones that are necessary here read
(in the $\overline{\rm MS}$ scheme with $D=4-2\e$)
\mathindent-2mm
\be \label{renor.const}
\begin{array}{rclrcl}
Z_{17} &=& -\fm{58}{243\e} \al^2 + {\cal O}(\al^3), &
Z_{18} &=& \fm{167}{648\e} \al^2 + {\cal O}(\al^3),\\[2mm]
Z_{27} &=& \fm{116}{81\e} \al^2 +\left( -\fm{23848}{2187\e^2} 
            + \fm{13390}{2187\e} \right) \al^3 + {\cal O}(\al^4), &
Z_{28} &=& \fm{19}{27\e} \al^2 + \left( -\fm{7249}{1458\e^2} 
            + \fm{5749}{5832\e} \right) \al^3 + {\cal O}(\al^4),\\[2mm]
Z_{47} &=& -\fm{50}{243\e} \al^2 + {\cal O}(\al^3), &
Z_{48} &=& -\fm{1409}{648\e} \al^2 + {\cal O}(\al^3),\\[2mm]
&& \hspace{-17mm} 
Z_{(11)7} =~ \fm{1096}{243} \al^2 + {\cal O}(\al^3), & 
&& \hspace{-17mm} 
Z_{(11)8} =~ -\fm{761}{162} \al^2 + {\cal O}(\al^3),\\[2mm]
Z_{77} &=& 1 -\fm{7}{3\e} \al + \left( \fm{35}{3\e^2} 
             + \fm{650}{27\e} \right) \al^2 + {\cal O}(\al^3), &
Z_{78} &=& 0,\\[2mm]
Z_{87} &=& -\fm{16}{9\e} \al + \left( \fm{104}{9\e^2} 
         -\fm{548}{81\e} \right) \al^2 + {\cal O}(\al^3), &
Z_{88} &=& 1 -\fm{3}{\e}\al + \left( \fm{16}{\e^2} 
         + \fm{1975}{108\e} \right) \al^2 + {\cal O}(\al^3).
\end{array} \hspace{-10mm}
\ee
\mathindent1cm
Their overall signs correspond to the following sign convention inside
the covariant derivative acting on a quark field $\psi$:
\be
D_{\mu} \psi = \left( \partial_{\mu} + i g G_{\mu}^a T^a + i e Q_{\psi}
A_{\mu} \right) \psi.
\ee
For completeness, one should also mention the $\overline{\rm MS}$
renormalization constant for the QCD gauge coupling in the
five-flavour effective theory ($g_{\rm bare} = Z_g g$)
\be \label{gauge.renor.const}
Z_g = 1 -\fm{23}{6\e} \al + \left( \fm{529}{24\e^2} 
                - \fm{29}{3\e} \right) \al^2 + {\cal O}(\al^3). 
\ee
Following Ref.~\cite{Bobeth:1999mk}, we ignore the quark-mass and
wave-function renormalization constants in the effective theory\footnote{
although their non-vanishing values were relevant in the calculations
of $Z_{ij}$ (\ref{renor.const}) and $Z_g$ (\ref{gauge.renor.const})}
because their effects cancel anyway in the matching condition with
analogous contributions on the full SM side. Only the top-quark
contributions to these renormalization constants will be included in
the SM counterterms (see Section~\ref{sec:SMren}).

The coefficients $C_k^{t(n)}$ vanish for $k=1,2,11$.  At the
tree-level, only $C^{c(0)}(\mu_0)=-1$ is different from zero. All the
$C^{Q(1)}_i(\mu_0)$ and $C^{Q(2)}_i(\mu_0)$ were found in
Ref.~\cite{Bobeth:1999mk} up to ${\cal O}(\e)$ and ${\cal O}(1)$,
respectively. In particular, $C_2^{c(1)}(\mu_0) = 0$ and
\bea
C_1^{c(1)}(\mu_0) &=& -15 - 6 \lw
   + \e \left( -\f{39}{2} - \f{\pi^2}{2} - 15 \lw - 3 \lwsq \right) + {\cal O}(\e^2),\\
C_4^{c(1)}(\mu_0) &=& \f{7}{9} - \f{2}{3} \lw    + \e \left( \f{77}{54} -\f{\pi^2}{18} 
                          + \f{7}{9} \lw - \f{1}{3} \lwsq \right) + {\cal O}(\e^2),\\
C_{11}^{c(1)}(\mu_0) &=& -\f{3}{2} - \lw + {\cal O}(\e),\\[2mm]
C_4^{t(1)}(\mu_0) &=& \left( 1 + \e \ln \f{\mu_0^2}{m_t^2} \right)
\left( \f{ -9 x^2 + 16 x - 4 }{ 6 (x - 1)^4} ~~\f{x^\e-1}{\e}
~+~ \f{7 x^3 + 21 x^2 - 42 x - 4}{36 (x - 1)^3} \right) \nnb\\[2mm]
&& +~ \e \left( \f{ -45 x^2 + 38 x + 28}{36 (x - 1)^4} \ln x
~+~ \f{23 x^3 + 93 x^2 + 66 x - 308}{216 (x - 1)^3} \right) + {\cal O}(\e^2),
\eea
where
\begin{eqnarray}
  x &=& \f{m_t^2(\mu_0)}{M_W^2}\,
\end{eqnarray}
has been introduced. For later convenience we also define the variables
\begin{eqnarray}
  w = \left( 1-\frac{M_W^2}{m_t^2(\mu_0)} \right)\,, \hspace{2cm}
  z = \frac{M_W^2}{m_t^2(\mu_0)}\,, \hspace{2cm}
  y = \frac{M_W}{m_t(\mu_0)} \,.
\end{eqnarray}
In the following, the $\overline{\rm MS}$-renormalized top-quark mass
$m_t(\mu_0)$ will often be denoted by just $m_t$.

For our present purpose, $C^{Q(1)}_{7,8}$ and $C^{Q(2)}_{7,8}$ are
needed up to ${\cal O}(\e^2)$ and ${\cal O}(\e)$, respectively. In
practice, this implies a necessity or repeating the one- and two-loop
matching computations for these coefficients from scratch. We shall
describe this calculation together with the three-loop one in the
following three sections.

\newsection{The unrenormalized SM amplitudes \label{sec:SMbare}}

\begin{figure}[t]
\vspace{5mm}
\hspace*{11mm}  $\gamma$ \hspace{36.5mm} $\gamma$ 
\hspace{41.5mm} $\gamma$ \hspace{36.5mm} $\gamma$ \\[7mm] 
\hspace*{6mm} $Q$ \hspace{9mm}     $Q$ \hspace{18mm} 
            $W^{\pm}$ \hspace{8mm}   $W^{\pm}$ \hspace{21mm}
              $Q$ \hspace{9mm}     $Q$ \hspace{20.5mm} 
          $\phi^{\pm}$ \hspace{8mm} $\phi^{\pm}$ \\[-18mm] 
\includegraphics[width=75mm,angle=0]{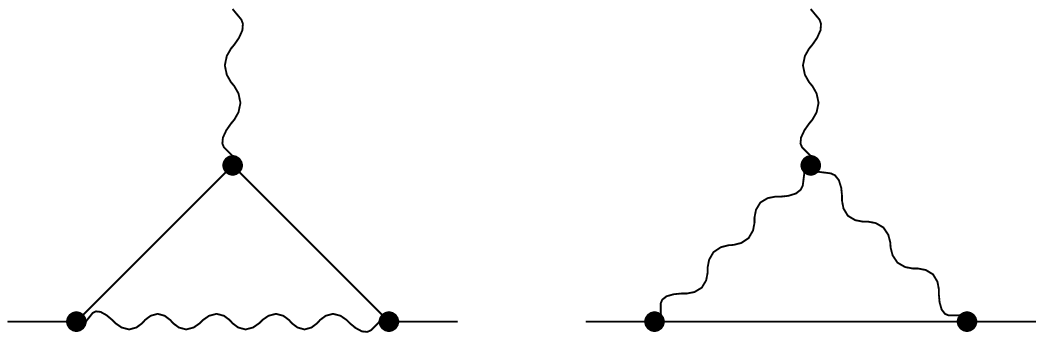}
\hspace{1cm}
\includegraphics[width=75mm,angle=0]{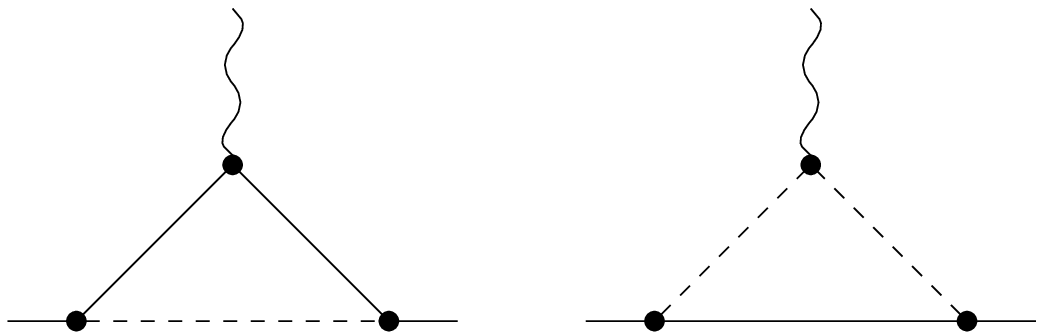}\\[-2mm]
$b$ \hspace{1cm}  $W^{\pm}$  \hspace{7mm} $s$ \hspace{6mm} 
$b$ \hspace{1cm}   $Q$   \hspace{1cm} $s$ \hspace{11mm} 
$b$ \hspace{1cm} $\phi^{\pm}$ \hspace{8mm} $s$ \hspace{6.5mm} 
$b$ \hspace{1cm}   $Q$   \hspace{1cm} $s$ \\[-1cm]
\begin{center}
\caption{\sf One-loop 1PI diagrams for $b \to s \gamma$ in the SM.
    There is no $W^{\pm}\phi^{\mp}\gamma$ coupling in the background field gauge.}
\label{fig:1loop.bsgamma}
\end{center}
\vspace{-1cm}
\end{figure}

We have to consider all the one-, two- and three-loop
one-particle-irreducible (1PI) diagrams contributing to the processes
$b\to s\gamma$ and $b\to sg$. The one-loop $b\to s\gamma$ diagrams
are shown in Fig.~\ref{fig:1loop.bsgamma}.  Higher-order diagrams are
found by adding internal gluons together with loop corrections on
their propagators.

We use the 't~Hooft-Feynman version of the background field gauge for
the electroweak interactions and QCD. Before performing the loop
integration, the Feynman integrands are Taylor-expanded up to second
order in the (off-shell) external momenta, and to the first order in
the $b$-quark mass.  Thus, effectively, the only massive particles in
our calculation are the top quark, the $W$ boson and the charged
pseudogoldstone scalar $\phi$.
The amputated 1PI\linebreak $b\to s\gamma$~ Green function can be cast into the
following form:
\begin{eqnarray}
  i \frac{4 G_F}{\sqrt{2}} \frac{eP_R}{(4 \pi)^2}
  \left[ V_{cs}^* V_{cb} 
    \sum_{j=1}^{13} Y^{c}_j(x) S_j
    + V_{ts}^* V_{tb} \sum_{j=1}^{13} Y^{t}_j(x) S_j 
  \right]
  \,,
\end{eqnarray}
with $P_R = (1+\gamma_5)/2$,
\bea \label{yc}
  Y^{c}_j(x) &=& \sum_{n\ge1} 
  A_c^{n\epsilon}\, \al^{n-1}\, Y_j^{c(n)}(x)
  \,,\\
  Y^{t}_j(x) &=& \sum_{n\ge1} 
  A_t^{n\epsilon}\, \al^{n-1}\, Y_j^{t(n)}(x)
  \,, \label{yt}
\eea
$A_c = \f{4\pi\mu_0^2}{M_W^2} e^{-\gamma}$ 
and 
$A_t = \f{4\pi\mu_0^2}{m_t^2} e^{-\gamma}$, where $\gamma$ is the
Euler constant. The symbols $S_j$ stand for different Dirac
structures that depend on the incoming $b$-quark momentum $p$ and on
the outgoing photon momentum $k$
\bea
  S_j &=& \left( \gamma_{\mu} \psla \ksla, \; 
    \gamma_{\mu} \; (p \cdot k), \;
    \gamma_{\mu} p^2, \;
    \gamma_{\mu} k^2, \;
    \psla k_{\mu}, \;
    \psla p_{\mu}, \;
    \ksla p_{\mu}, \;
    \ksla k_{\mu}, \;
  \right. \nonumber\\ && \hspace{4cm} \left. \label{DirStructures} 
    m_b \ksla \gamma_{\mu}, \;
    m_b \gamma_{\mu} \ksla, \;
    m_b \psla \gamma_{\mu}, \;
    m_b \gamma_{\mu} \psla, \;
    M_W^2 \gamma_{\mu} \right)_j.
\eea
The first two terms in the expansion of ~$Y^c_j$~(\ref{yc}) are
$x$-independent, but the third (three-loop) and higher terms do depend
on $x$.

By analogy, the ~$b\to s g$~ Green function reads
\begin{eqnarray}
  i \frac{4 G_F}{\sqrt{2}} \frac{g P_R T^a}{(4 \pi)^2}
  \left\{ V_{cs}^* V_{cb}  
    \sum_{j=1}^{13} G^{c}_j(x) S_j
    + V_{ts}^* V_{tb} \sum_{j=1}^{13} G^{t}_j(x) S_j 
  \right\}
  \,,
\end{eqnarray}
with
\begin{eqnarray}
  G^{c}_j(x) &=& \sum_{n\ge1} 
  A_c^{n\epsilon}\, \al^{n-1}\, G_j^{c(n)}(x)
  \,,\\
  G^{t}_j(x) &=& \sum_{n\ge1} 
  A_t^{n\epsilon}\, \al^{n-1}\, G_j^{t(n)}(x)
  \,.
\end{eqnarray}

As shown in Refs.~\cite{Grinstein:tj,Bobeth:1999mk}, only the
following linear combinations of ~$Y_j^{Q(n)}$ and $G_j^{Q(n)}$ are
sufficient for finding the coefficients $C_7(\mu_0)$ and $C_8(\mu_0)$:
\bea
C_{7,{\rm bare}}^{Q(n)} &\equiv& \f{1}{4} Y_2^{Q(n)} +  Y_{10}^{Q(n)}, \label{cb7}\\
C_{8,{\rm bare}}^{Q(n)} &\equiv& \f{1}{4} G_2^{Q(n)} +  G_{10}^{Q(n)}. \label{cb8}
\eea

The calculation of $C_{k,{\rm bare}}^{Q(2)}$ up to ${\cal O}(\e)$
requires supplementing Eqs.~(57) and (58) of Ref.~\cite{Bobeth:1999mk}
by higher orders in $\e$, which yields\footnote{
All the other equations in Section 5.1 of Ref.~\cite{Bobeth:1999mk}
are valid to all orders in $\e$.}
\mathindent0cm
\be
I^{(2)}_{n_1 n_2 n_3} \bbuildrel{=\!=\!=\!=}_{\scs m_2 = 0}^{}
(-1)^{N+1} 
\f{ (1+2\e)_{N-5} (1+\e)_{n_2+n_3-3} (1-\e)_{1-n_2} (1-\e)_{1-n_3}}{ 
(n_1-1)! (n_2-1)! (n_3-1)! (1-\e)} \f{\Gamma(1+2\e) \Gamma(1-\e)}{\Gamma(1+\e)}
\ee
and
\bea
I^{(2)}_{111} &=& 
\f{1}{2 (1-\e) (1-2\e)} \left\{ -\f{1+r}{\e^2} \;+\; \f{2}{\e} r \ln r \;+\;
(1-2r) \ln^2 r \;+\; 2 (1-r) \mbox{Li}_2\left(1-\f{1}{r}\right) 
\right. \nnb\\[2mm] && \left. \hspace{-14mm}
+ 2 \e (1-r) \left[ \mbox{Li}_3\left(1-r\right) - \mbox{Li}_3\left(1-\f{1}{r}\right) 
                       - \mbox{Li}_2\left(1-\f{1}{r}\right) \ln r \right]
   + \e \left(r-\f{2}{3}\right) \ln^3 r + {\cal O}(\e^2) \right\}\!\!,
\eea
\mathindent1cm
for the generic two-loop integral
\be \label{int2}
I^{(2)}_{n_1 n_2 n_3} = 
\f{(m_1^2)^{N-4+2\e}}{\pi^{4-2\e}\; \Gamma(1+\e)^2}
\int \f{d^{4-2\e}q_1 \; d^{4-2\e}q_2}{(q_1^2 - m_1^2)^{n_1}
                  (q_2^2 - m_2^2)^{n_2}[(q_1 - q_2)^2]^{n_3}},
\ee
where $r = m_2^2/m_1^2$, $N=n_1+n_2+n_3$ and $(a)_n =
\Gamma(a+n)/\Gamma(n)$. Otherwise, the calculation proceeds precisely
as described in Section~5 of that paper. The unrenormalized one- and
two-loop results read
\mathindent0cm
\bea
C_{7,{\rm bare}}^{c(1)} &=& \fm{23}{36} + \fm{145\e}{216} +\fm{875\e^2}{1296} 
                             + \fm{23\e^2\pi^2}{432} + {\cal O}(\e^3),\\[2mm]
C_{8,{\rm bare}}^{c(1)} &=& \fm{1}{3} + \fm{11\e}{18} + \fm{85\e^2}{108} 
                                 + \fm{\e^2\pi^2}{36} + {\cal O}(\e^3),\\[2mm]
C_{7,{\rm bare}}^{c(2)} &=& \fm{112}{81\e} -\fm{107}{243} 
                  -\fm{4147\e}{1458} +\fm{56\e\pi^2}{81} + {\cal O}(\e^2),\\[2mm]
C_{8,{\rm bare}}^{c(2)} &=& \fm{23}{27\e} +\fm{833}{324} 
              + \fm{13429\e}{1944} + \fm{23\e\pi^2}{54}  + {\cal O}(\e^2),\\[2mm]
C_{7,{\rm bare}}^{t(1)} &=& \left( 1 + \fm{\e^2 \pi^2}{12} \right)  
\left( \fm{3 x^3 - 2 x^2}{4 (x - 1)^4} ~\fm{x^\e - 1}{\e} 
     + \fm{22 x^3 - 153 x^2 + 159 x - 46}{72 (x - 1)^3} \right)
\nnb\\ && 
+ \e \left( \fm{-18 x^3 + 150 x^2 - 157 x + 46}{72 (x - 1)^4} ~\fm{x^\e - 1}{\e} 
       + \fm{122 x^3 - 933 x^2 + 975 x - 290}{432 (x - 1)^3} \right)
\nnb\\ && 
+ \e^2 \left( \fm{-108 x^3 + 918 x^2 - 977 x + 290}{432 (x - 1)^4} \ln x
 + \fm{694 x^3 - 5619 x^2 + 5937 x - 1750}{2592 (x - 1)^3} \right) + {\cal O}(\e^3),\\[2mm]
C_{8,{\rm bare}}^{t(1)} &=& \left( 1 + \fm{\e^2 \pi^2}{12} \right)  
\left(   \fm{-3 x^2}{4 (x - 1)^4} ~\fm{x^\e - 1}{\e} 
       + \fm{5 x^3 - 9 x^2 + 30 x - 8}{24 (x - 1)^3} \right)
\nnb\\ && 
+ \e \left(   \fm{-15 x^2 - 14 x + 8}{24 (x - 1)^4} ~\fm{x^\e - 1}{\e} 
       + \fm{13 x^3 + 15 x^2 + 186 x - 88}{144 (x - 1)^3} \right)
\nnb\\ && 
+ \e^2 \left( \fm{-81 x^2 - 130 x + 88}{144 (x - 1)^4}  \ln x
+ \fm{35 x^3 + 273 x^2 + 1110 x - 680}{864 (x - 1)^3} \right) + {\cal O}(\e^3),\\[2mm]
C_{7,{\rm bare}}^{t(2)} &=& \fm{1}{\e} \left( 1 + \fm{\e^2\pi^2}{6} \right) 
\left(   \fm{-6 x^4 - 46 x^3 + 28 x^2}{3 (x - 1)^5} \fm{x^{\e} - 1}{\e}
       + \fm{34 x^4 + 101 x^3 + 402 x^2 - 397 x + 76}{27 (x - 1)^4}  \right)
\nnb\\ && 
+ \fm{-16 x^4 - 122 x^3 + 80 x^2 - 8 x}{9 (x - 1)^4} H(x,\e)
+ \fm{-333 x^4 - 2529 x^3 + 688 x^2 + 778 x - 224}{81 (x - 1)^5} \fm{x^{\e} - 1}{\e}
\nnb\\ && 
+ \fm{-220 x^4 + 12952 x^3 - 9882 x^2 + 2407 x - 397}{243 (x - 1)^4} 
+ \e \left[ 
\fm{146 x^4 - 4289 x^3 + 2736 x^2 + 14 x - 224}{81 (x - 1)^4} \mbox{Li}_2\left(1-\f{1}{x}\right) 
\right. \nnb\\ && \left. \hspace{-8mm}
 + \fm{-879 x^4 - 50319 x^3 + 35810 x^2 - 5884 x + 428}{486 (x - 1)^5}  \ln x
 + \fm{-4381 x^4 + 148252 x^3 - 89391 x^2 + 8797 x - 745}{1458 (x - 1)^4} \right]
+ {\cal O}(\e^2),\\[2mm]
C_{8,{\rm bare}}^{t(2)} &=& \fm{1}{\e} \left( 1 + \fm{\e^2\pi^2}{6} \right) 
\left(
    \fm{17 x^3 + 31 x^2}{2 (x - 1)^5} \fm{x^{\e} - 1}{\e}
  + \fm{35 x^4 - 170 x^3 - 447 x^2 - 338 x + 56}{36 (x - 1)^4} \right)
\nnb\\ && 
+ \fm{-4 x^4 + 40 x^3 + 41 x^2 + x}{6 (x - 1)^4} H(x,\e)
+ \fm{-144 x^4 + 4707 x^3 + 8887 x^2 - 122 x - 368}{216 (x - 1)^5} \fm{x^{\e} - 1}{\e}
\nnb\\ && 
+ \fm{-1367 x^4 - 9646 x^3 - 76869 x^2 + 7442 x + 2680}{1296 (x - 1)^4} 
+ \e \left[
  \fm{641 x^4 + 184 x^3 + 8001 x^2 - 220 x - 368}{216 (x - 1)^4} \mbox{Li}_2\left(1-\f{1}{x}\right) 
\right. \nnb\\ && \left. \hspace{-16mm}
+ \fm{2982 x^4 + 30843 x^3 + 147437 x^2 - 7846 x - 6664}{1296 (x - 1)^5}  \ln x
+ \fm{-22703 x^4 - 56674 x^3 - 934701 x^2 - 46090 x + 59656}{7776 (x - 1)^4} \right]
+ {\cal O}(\e^2),
\eea
\mathindent1cm
where
\bea
H(x,\e) &=& \mbox{Li}_2\left(1-\fm{1}{x}\right) 
         + \e \left[ \mbox{Li}_3\left(1-x\right) - \mbox{Li}_3\left(1-\fm{1}{x}\right)
        + \mbox{Li}_2\left(1-\fm{1}{x}\right) \ln x  + \fm{1}{6} \ln^3 x \right].
\eea

In addition to the bare coefficients, we shall also need those parts
of $C_{i,{\rm bare}}^{t(1)}$ that originate from the $m_b$-dependent
Dirac structure $S_{10}$, as they play a separate role when $m_b$
gets renormalized. They read
\bea \label{bb7}
B_7 &\equiv& Y_{10}^{t(1)} = \left( 1 + \fm{\e^2\pi^2}{12} \right) 
\left( \fm{-3 x^2 + 2 x}{6 (x - 1)^3} \; \fm{x^{\e}-1}{\e} 
     + \fm{5 x^2 - 3 x}{12 (x - 1)^2} \right)
+ \e \left( \fm{-2 x^2 + x}{4 (x - 1)^3} \; \fm{x^{\e}-1}{\e} 
          + \fm{11 x^2 - 5 x}{24 (x - 1)^2} \right)
\nnb\\[1mm] && \hspace{13mm} 
+ \e^2 \left( \fm{-12 x^2 + 5 x}{24 (x - 1)^3} \ln x
            + \fm{23 x^2 - 9 x}{48 (x - 1)^2} \right) 
+ {\cal O}(\e^3),\\[2mm]
B_8 &\equiv& G_{10}^{t(1)} = \left( 1 + \fm{\e^2\pi^2}{12} \right) 
\left( \fm{x}{2 (x - 1)^3} \; \fm{x^{\e}-1}{\e} 
     + \fm{x^2 - 3 x}{4 (x - 1)^2} \right)
+ \e \left( \fm{3 x}{4 (x - 1)^3} \; \fm{x^{\e}-1}{\e} 
          + \fm{x^2 - 7 x}{8 (x - 1)^2} \right) 
\nnb\\[1mm] && \hspace{13mm} 
+ \e^2 \left( \fm{7 x}{8 (x - 1)^3} \ln x
            + \fm{x^2 - 15 x}{16 (x - 1)^2} \right)
+ {\cal O}(\e^3). \label{bb8}
\eea
The renormalization of $m_b$ will not matter in the charm sector because
$Y_{10}^{c(1)} = G_{10}^{c(1)} = 0$.\\

Let us now turn to the main purpose of our paper, i.e. to the
three-loop calculation. One of the ${\cal O}(10^3)$ diagrams that we
have calculated at this level is shown in Fig.~\ref{fig:sampleSM}.
Obviously, when the virtual top quark is present in the open fermion
line, we have to deal with three-loop vacuum integrals involving two
mass scales, $m_t$ and $M_W$. However, such double-scale integrals are
encountered in the charm-quark sector, too, when closed top-quark
loops arise on the virtual gluon lines.

At present, complete three-loop algorithms exist for vacuum integrals
involving only a single mass scale. We have reduced our calculation to
such integrals by performing expansions around the point $m_t=M_W$ and
for $m_t \gg M_W$. In the latter case, the method of asymptotic
expansions of Feynman integrals has been applied \cite{Smi02}. At the
physical point where $M_W/m_t\approx0.5$, both expansions work
reasonably well (see Section~\ref{sec:results}).

Two different approaches have been used for the calculation of the
three-loop diagrams.  The first one is based on a completely automated
set-up where the diagrams are generated by {\tt
QGRAF}~\cite{Nogueira:1993ex}, further processed with {\tt
q2e}~\cite{q2e} and {\tt exp}~\cite{exp}, and finally evaluated and
expanded in $\epsilon$ with the help of the package {\tt
MATAD}~\cite{Steinhauser:2000ry} written in {\tt Form}~\cite{Form}.
{\tt MATAD} is designed to compute single-scale vacuum integrals up to
three loops.  The individual packages work hand in hand, and thus no
additional manipulation from outside is necessary. Moreover, all the
auxiliary files, e.g. make-files to control the calculation or files
to sum the individual diagrams, are generated automatically.

The program {\tt exp} is designed to automatically apply the rules of
asymptotic expansions in the limit of large external momenta or
masses.  Thus, its output crucially depends on the limit we consider.
For the expansion around $m_t = M_W$, the asymptotic expansion reduces
to the usual Taylor expansion in powers of $w \equiv (1-M_W^2/m_t^2)$
and thus {\tt exp} essentially rewrites the output of {\tt q2e} to a
format suitable for {\tt MATAD}.  However, for $m_t\gg M_W$, next to
the Taylor expansion in $z = M_W^2/m_t^2$, more diagrams expanded in
various small quantities contribute according to the rules of
asymptotic expansions. The package {\tt exp} provides a proper input
for {\tt MATAD} which then performs the expansions up to the required
depth, and computes the resulting scalar vacuum integrals. The mass
scale of the latter is either given by $m_t$ or $M_W$.

An important element of the calculation are the so-called projection
operations that pick only the two Dirac structures we need (see
Eqs.~(\ref{cb7}) and~(\ref{cb8})), and thus bypass the time-consuming
tensor algebra.

Using this method, we evaluated the expansions in $z$ and $w$ up to
orders $z^4$ and $w^6$, respectively. Furthermore, it was possible to
compute the first few expansion terms for general gauge parameter
$\xi$, in order to check that it drops out in the sum of all bare
three-loop diagrams. This imposes a strong check on the correctness of
our results.

In the second approach,~ {\tt MATAD} was also used for three-loop
scalar integrals involving a single mass scale. However, the diagrams
were generated using {\tt FeynArts} \cite{Kublbeck:xc}. The remaining
part of the calculation was performed with the help of self-written
programs, largely overlapping with those used several years ago for
the calculation of three-loop anomalous dimension matrices
\cite{Chetyrkin:1996vx}.  No projection operations were used, and all
the Dirac structures (except for the ones quadratic in $k$) appeared
in the results, which allowed for performing several consistency
checks. This approach was obviously much slower, and was finally
brought through thanks to the use of the {\tt Z-Box} computer\footnote{
{\tt http://krone.physik.unizh.ch/$\sim$stadel/zBox}}
at the University of Z\"urich. Only the expansion around
$m_t=M_W$ (up to $w^8$) was calculated using this method.

Although our results for the three-loop diagrams are known in terms of
expansions only, we are able to determine the exact $x$-dependence of
their pole parts by using the matching equation discussed in
Section~\ref{sec:matching}. Of course, we have verified that these
pole parts have precisely the same expansions in $z$ and $w$
as found from the direct calculation up to $z^4$ and $w^8$.

Our results for $C_{7,{\rm bare}}^{Q(3)}$ and $C_{8,{\rm bare}}^{Q(3)}$
take the following form:
\bea
C_{7,{\rm bare}}^{c(3)} &=& \f{10798}{2187\,\e^2} + \f{1}{\e} \left[ - \f{215}{162} 
                             - \f{8}{405\,x} + \f{224}{243} \ln x \right] + f_7^c(x),\\[2mm]
C_{8,{\rm bare}}^{c(3)} &=& \f{4675}{1458\,\e^2} + \f{1}{\e} \left[ \f{11783}{648} 
                             - \f{169}{2160\,x} + \f{46}{81} \ln x \right] + f_8^c(x),\\[2mm]
C_{7,{\rm bare}}^{t(3)} &=& p_7^t(x) + f_7^t(x),\\[2mm]
C_{8,{\rm bare}}^{t(3)} &=& p_8^t(x) + f_8^t(x),
\eea
where the pole parts in the top sector read
\bea
p_7^t(x) &=& \fm{1}{\e^2} \left[ 
 \fm{-57 x^5 + 634 x^4 + 1911 x^3 - 1044 x^2 - 4 x}{9 (x - 1)^6} \; 
\left( \ln x + \fm{\e}{2} \ln^2 x \right)
\right.\nnb\\[1mm] && \left.   
+\fm{380 x^5 - 1099 x^4 - 8521 x^3 - 4385 x^2 + 5797 x - 812}{54 (x - 1)^5} \right] 
\nnb\\[1mm] && 
+ \fm{1}{\e} \left[ 
 \fm{-560 x^5 + 190 x^4 + 12410 x^3 - 6200 x^2 + 496 x}{27 (x - 1)^5} 
         \mbox{Li}_2\left(1-\f{1}{x}\right)
\right.\nnb\\[1mm] && \left.   
+\fm{-35586 x^5 + 223524 x^4 + 1345261 x^3 - 604386 x^2 
                 - 55425 x + 20852}{1458 (x - 1)^6} \ln x 
\right.\nnb\\[1mm] && \left.   
+\fm{-325132 x^6 + 7070681 x^5 - 72622435 x^4 + 48723685 x^3 
          - 11218745 x^2 + 1543882 x + 864}{43740\, x (x - 1)^5} 
\right], \\[3mm]
p_8^t(x) &=& \fm{1}{\e^2} \left[
 \fm{-199 x^4 - 3018 x^3 - 2535 x^2 - 8 x}{12 (x - 1)^6} \; 
\left( \ln x + \fm{\e}{2} \ln^2 x \right)
\right.\nnb\\[1mm] && \left.   
+\fm{2054 x^5 - 11080 x^4 + 52535 x^3 + 105505 x^2 + 26875 x - 3089}{360 (x - 1)^5} \right]
\nnb\\[1mm] && 
+ \fm{1}{\e} \left[ 
 \fm{-140 x^5 + 964 x^4 - 4813 x^3 - 3440 x^2 - 59 x}{18 (x - 1)^5} 
         \mbox{Li}_2\left(1-\f{1}{x}\right)
\right.\nnb\\[1mm] && \left.  
+\fm{-75843 x^5 - 11835 x^4 - 9946790 x^3 - 8078850 x^2 
               + 114225 x + 114293}{9720 (x - 1)^6} \ln x 
\right.\nnb\\[1mm] && \left.   
+\fm{-5561837 x^6 + 4392955 x^5 + 397608280 x^4 + 760910570 x^3 - 
          79703785 x^2 - 4603813 x + 45630}{583200\, x (x - 1)^5} 
\right]. 
\eea
For the UV- and IR-finite functions $f_k^Q(x)$, we write the
expansions as follows:
\begin{eqnarray}
  f_k^Q(x) &=& \sum_{n,m} a_{nm}^{kQ} \;\; \f{\ln^m x}{x^n} ~~\equiv~~
   \sum_{n,m} a_{nm}^{kQ} \;\; (-1)^m z^n \ln^m z \,, 
   \hspace{1cm} (m \leq 3),   \label{exp.f.a}\\
  f_k^Q(x) &=& \sum_{n} b_n^{kQ} \left(1-\f{1}{x} \right)^n ~~\equiv~~
    \sum_{n} b_n^{kQ} \; w^n \,.  \label{exp.f.b}
\end{eqnarray}
The values of $a_{nm}^{kQ}$ and $b_n^{kQ}$ that we have found are
listed in Appendix~A.

\newsection{The SM counterterms \label{sec:SMren}}

The renormalization scheme that we apply on the SM side is chosen in
such a way that values of the renormalized $\alpha_s$, the light-quark
wave-functions and masses overlap with their $\overline{\rm MS}$
counterparts in the five-flavour effective theory. Thus, $\alpha_s$
means $\alpha_s^{(5)}(\mu_0)$ throughout the paper. The one-loop
renormalization constant of the QCD gauge coupling in the SM reads
(cf.~Eq.~(\ref{gauge.renor.const}))
\be \label{SM.gauge.renor.const}
Z_g^{\rm SM} = 1+ \f{\al}{\e} \left( -\f{23}{6} + \f{1}{3} N_{\e} \right) + {\cal O}(\al^2).
\ee
Here, $N_{\e}$ parametrizes the one-loop threshold correction that
arises in the relation between $\alpha_s^{(6)}$ and $\alpha_s^{(5)}$,
i.e. when the top quark is decoupled from $\alpha_s$. A collection of
explicit expressions for such corrections (also called ``decoupling
constants'') up to three loops can be found in
Ref.~\cite{Steinhauser:2002rq}. The value 
\be
N_{\e} = \left( \f{4\pi\mu_0^2}{m_t^2} \right)^{\!\displaystyle \e}\; \Gamma(1+\e)
\ee
is found (exactly in $\e$) from the requirement that the top-quark
loop contribution to the off-shell background gluon propagator with
momentum $q$ is renormalized away, up to effects of order
~$q^2/m_t^2$~ that match onto higher-dimensional operators in the
effective theory.

The same requirement applied to the light-quark propagators at two
loops leads to the following expressions for the renormalization
constants of their wave-functions and masses 
($\psi_{\rm bare} = Z_{\psi} \psi$,~~$m_{\rm bare} = Z_m m$)
\bea 
\Delta Z_{\psi} &\equiv& \label{ZdifPsi}
Z_{\psi}^{\rm SM} - Z_{\psi}^{\rm eff.~theory} ~~=~~ 
\al^2 N_{\e}^2 \left( \f{2}{3\e} - \f{5}{9} \right) + {\cal O}(\al^3,\e),\\
\Delta Z_m &\equiv& \label{ZdifM}
Z_m^{\rm SM} - Z_m^{\rm eff.~theory} ~~=~~ 
\al^2 N_{\e}^2 \left( -\f{4}{3\e^2} + \f{10}{9\e} -\f{89}{27} \right) + {\cal O}(\al^3,\e).
\eea
In our calculation, the latter renormalization constant matters for
the $b$-quark only, because we include linear terms in $m_b$, while
all the other light particles are treated as massless.

The differences $\Delta Z_{\psi}$ and $\Delta Z_m$ are everything we
need to know about the renormalization of the light-quark
wave functions and masses. Since the wave-function renormalization
matters for external fields only, the remaining parts of the
considered renormalization constants cancel out in the matching
equation, i.e. in the difference between the full SM and the effective
theory off-shell amplitudes. It is worth noticing that since $\Delta
Z_{\psi}$ and $\Delta Z_m$ arise at ${\cal O}(\al^2)$ only, they had
no effect on the two-loop matching computation in
Ref.~\cite{Bobeth:1999mk}.

As far as the top-quark mass is concerned, we renormalize it in the
$\overline{\rm MS}$ scheme, at the scale $\mu_0$, in the six-flavour
QCD. The corresponding renormalization constant, when expressed in
terms of $\al \equiv \al^{(5)}(\mu_0)$, takes the following form
(exactly in $\e$)
\mathindent5mm
\be
Z_{m_t} \equiv 1 + \al Z_{m_t}^{(1)} + \al^2 Z_{m_t}^{(2)} + {\cal O}(\al^3) ~=~ 
  1 -\f{4}{\e} \al + \left( \f{74}{3\e^2} - \f{27}{\e} 
- \f{8}{3\e^2} N_{\e} \right) \al^2 + {\cal O}(\al^3).
\ee
\mathindent1cm

Two more QCD renormalization constants need to be thought about in the
context of our calculation. The first of them is the external gluon
wave-function renormalization constant in the $b \to sg$ case.  In the
background field gauge, it just cancels with the renormalization of
the gauge coupling in the vertex where the external gluon is
emitted.\footnote{
In the usual (non-background) 't~Hooft-Feynman gauge, we would need to
introduce, by analogy to Eqs.~(\ref{ZdifPsi}) and (\ref{ZdifM}),
$\Delta \left( Z_g \sqrt{Z_G} \right) = \al^2 N_{\e}^2 \left(
-3/(4\e^2) + 5/(8\e) - 89/48 \right) + {\cal O}(\al^3,\e)$.}
The second one is the renormalization constant of the QCD gauge-fixing
parameter $\xi$. It plays no role either, because $C_{7,{\rm
bare}}^{Q(2)}$ and $C_{8,{\rm bare}}^{Q(2)}$ are
$\xi$-independent.\footnote{
Contrary to the bare two-loop Wilson coefficients of the EOM-vanishing
operators (Eq.~(73) of Ref.~\cite{Bobeth:1999mk}).}

Last but not least, one needs to consider possible electroweak
counterterms. Since we work at the leading order in the electroweak
interactions, the only electroweak counterterms that may matter for us
must have the $\bar{s}b$ flavour content. Their dimensionality cannot
exceed 4, and they must be invariant under the QCD and QED gauge
transformations. These conditions leave out only two possible
electroweak counterterms: $\bar{s} \Dsla b$ and $\bar{s} b$. They
originate from the flavour-off-diagonal renormalization of the quark
wave-functions and Yukawa matrices. Since we refrain from applying
unitarity of the CKM matrix (but set $V_{ub}$ to zero), we write the
corresponding electroweak counterterm Lagrangian as follows:
\mathindent0cm
\be \label{eq:Lew}
{\cal L}^{\rm ew}_{\rm counter} = \f{G_F}{\pi\sqrt{2}} 
\left[ 
V^*_{cs} V_{cb} A_c^{\e} \; \bar{s} \left( i Z_{2,sb}^c \Dsla - Z^c_{0,sb}\, m_b \right) b
\; + \; 
V^*_{ts} V_{tb} A_t^{\e} \; \bar{s} \left( i Z_{2,sb}^t \Dsla - Z^t_{0,sb}\, m_b \right) b
\right], 
\ee
\mathindent1cm
with the factors $A_c$ and $A_t$ that have been defined below Eq.~(\ref{yt}).
The renormalization constants $Z_{2,sb}^Q$ and $Z_{0,sb}^Q$ are fixed by
the requirement that the renormalized off-shell light-quark
propagators with momentum $q$ remain flavour-diagonal, up to effects
of order $q^2/M_W^2$ that match onto higher-dimensional operators in
the effective theory. A simple one-loop calculation gives
\bea
Z_{2,sb}^c &=& -\f{2-2\e}{2-\e}\, \Gamma(\e),\\
Z_{0,sb}^c &=& 0,\\
Z_{2,sb}^t &=& \Gamma(\e) \left[ -\f{x}{2} - 1 
+ \f{ 2 x^2 + 3 x - 2}{2 (x - 1)^2} (x^{\e}-1) + \e \f{-3 x^2 - x - 2}{4 (x - 1)}
\right. \nnb\\ && \left. \hspace{2cm}
+ \e^2 \left( \f{4 x^2 - x + 2}{4 (x - 1)^2} \ln x 
      + \f{-7 x^2 - x - 2}{8 (x - 1)} \right)  + {\cal O}(\e^3)\right],\\[2mm]
Z_{0,sb}^t &=& \f{x(x^{\e}-x)\Gamma(\e)}{(1-\e)(x-1)}.
\eea
Higher-order (in $\al$) contributions to $Z_{2,sb}^Q$ and $Z_{0,sb}^Q$
are irrelevant to us, because the counterterms (\ref{eq:Lew}) affect
our calculation only when inserted into two-loop diagrams containing
top-quark loops on the gluon lines. Otherwise, the loop integrals
vanish in dimensional regularization after expanding them in external
momenta, because all the propagator denominators are massless. As far
as the tree-level diagrams are concerned, they give no contribution to
the relevant structures $S_2$ and $S_{10}$ in Eq.~(\ref{DirStructures}).

\newpage
\newsection{Matching \label{sec:matching}}

We are now ready to write down the matching equation that follows from
the requirement of equality of the effective theory and the full SM
amputated 1PI Green functions. The former ones originate from
tree-level diagrams only, because all the loop integrals with massless
denominators vanish in dimensional regularization, after expanding them
in external momenta.

For the coefficients $C_i^Q$ ($i=7,8$), the matching equation up to
three loops takes the following form:
\mathindent0cm
\bea 
\left( Z_g^2 \al \right)^{-1} \sum_k C_k^{Q} Z_{ki} &=&
\left( 1 + \Delta Z_{\psi} \right) \sum_{n=1}^3 \al^{n-1} 
\left[ \left(Z_g^{\rm SM}\right)^{2(n-1)} A_Q^{n\e} C_{i,{\rm bare}}^{Q(n)} 
         + \delta^{tQ} T_i^{(n)} \right]
\nnb\\ && 
+ \delta^{tQ} A_t^{\e} \Delta Z_m B_i  
+ \fm{1}{x} A_t^{2\e} A_Q^{\e}\, \al^2 
\left( Z_{2,sb}^Q K_i + Z_{0,sb}^Q R_i \right) + {\cal O}(\al^3).
\label{main.matching}
\eea
\mathindent1cm
Non-vanishing contributions on the l.h.s. arise for $k=1,2,4,7,8,11$
that we have considered in Section~\ref{sec:eff.theory}. The effect
of $m_b$-renormalization is contained in the $\Delta Z_m B_i$ term, where
$B_i$ have been given in Eqs.~(\ref{bb7}) and (\ref{bb8}).

The quantities $T_i^{(n)}$ originate from the top-quark mass
renormalization. Replacing in the bare results $m_t$ by $Z_{m_t} m_t$
and Taylor-expanding in $\al$, one finds $T_i^{(1)} = 0$,
\bea
T_i^{(2)} &=& 2 A_t^{\e} Z_{m_t}^{(1)} \left(x\f{\partial}{\partial x} -\e \right) 
              C_{i,{\rm bare}}^{t(1)}\,,\\ 
T_i^{(3)} &=& 
A_t^{\e} \left[ 2 Z_{m_t}^{(2)} \left(x\f{\partial}{\partial x} -\e \right) 
    + \left( Z_{m_t}^{(1)} \right)^2  \left( 2 x^2\f{\partial^2}{\partial x^2}  
+ (1-4\e) x\f{\partial}{\partial x} +\e + 2\e^2 \right) \right] C_{i,{\rm bare}}^{t(1)}
\nnb\\ && 
+ 2 A_t^{2\e} Z_{m_t}^{(1)} \left( x\f{\partial}{\partial x} -2\e \right) C_{i,{\rm bare}}^{t(2)}\,.
\eea
The explicit factors of $\e$ in the above equation are due to the fact
that $A_t$ depends on $m_t$, too.

The quantities $K_i$ and $R_i$ on the r.h.s. of
Eq.~(\ref{main.matching}) originate from two-loop $b \to s \gamma$ and
$b \to sg$ diagrams with insertions of the electroweak counterterm
(\ref{eq:Lew}) and with closed top-quark loops on the gluon lines. 
We find
\be \label{sDb}
\begin{array}{rclcrcl}
K_7 &=& -\fm{8}{405} + \fm{88\,\e}{6075} + {\cal O}(\e^2), &~~&
K_8 &=& -\fm{169}{2160} + \fm{10333\,\e}{64800}  + {\cal O}(\e^2),\\[2mm]
R_7 &=& {\cal O}(\e^2), &&
R_8 &=& \fm{1}{40\e} +\fm{193}{1200} - \fm{8441\,\e}{36000} + \fm{\pi^2\e}{240} + {\cal O}(\e^2).\\
\end{array}
\ee
It is interesting to notice that the ~$m_b \bar{s} b$~ counterterm from
Eq.~(\ref{eq:Lew}) is irrelevant for $C_7^{t(3)}$ (because $R_7= {\cal
O}(\e^2)$) and for the charm sector (because $Z^c_{0,sb} =0$). Thus,
it matters for $C_8^{t(3)}$ only.

At this point, all the ingredients of the r.h.s. of the
Eq.~(\ref{main.matching}) have been explicitly specified. As far as
the l.h.s. of this equation is concerned, Section~\ref{sec:eff.theory}
provides us with all the necessary renormalization constants and
Wilson coefficients, except for $C^Q_7$ and $C^Q_8$. Thus, we can find
$C^{Q(n)}_7$ and $C^{Q(n)}_8$ for $n=1,2,3$ by solving our matching
equation (\ref{main.matching}) order-by-order in $\al$. All the
$1/\e^2$ and $1/\e$ poles cancel during this operation, as they
should. The resulting finite Wilson coefficients are presented in the
next section.

\newsection{Results \label{sec:results}}

Our final results for the renormalized Wilson coefficients of the
operators $P_7$ and $P_8$ are as follows:
\mathindent0cm
\bea
C_7^{c(1)}(\mu_0) &=& \fm{23}{36} 
+\e \left( \fm{145}{216} + \fm{23}{36} \ln \fm{\mu_0^2}{M_W^2} \right) 
+\e^2 \left( \fm{875}{1296} + \fm{23\pi^2}{432}
+ \fm{145}{216} \ln \fm{\mu_0^2}{M_W^2}  
+ \fm{23}{72} \ln^2 \fm{\mu_0^2}{M_W^2} \right) +\!{\cal O}(\e^3),\\[3mm] 
C_8^{c(1)}(\mu_0) &=& \fm{1}{3}
+\e \left( \fm{11}{18} + \fm{1}{3} \ln \fm{\mu_0^2}{M_W^2} \right)  
+\e^2 \left( \fm{85}{108} + \fm{\pi^2}{36}
+ \fm{11}{18 } \ln \fm{\mu_0^2}{M_W^2}  
+ \fm{1}{6} \ln^2 \fm{\mu_0^2}{M_W^2} \right) + {\cal O}(\e^3),\\[3mm] 
C_7^{c(2)}(\mu_0) &=& -\fm{713}{243} - \fm{4}{81} \ln \fm{\mu_0^2}{M_W^2}  
+ \e \left( -\fm{7357}{1458} +\fm{37\pi^2}{81}
- \fm{820}{243} \ln \fm{\mu_0^2}{M_W^2}  
+ \fm{110}{81} \ln^2 \fm{\mu_0^2}{M_W^2} \right) + {\cal O}(\e^2),\\[3mm] 
C_8^{c(2)}(\mu_0) &=&  -\fm{91}{324} + \fm{4}{27} \ln \fm{\mu_0^2}{M_W^2}  
+ \e \left( \fm{6289}{1944} + \fm{8\pi^2}{27}
+ \fm{371}{162} \ln \fm{\mu_0^2}{M_W^2}  
+ \fm{25}{27} \ln^2 \fm{\mu_0^2}{M_W^2} \right) + {\cal O}(\e^2),\\[3mm] 
C_7^{c(3)}(\mu_0) &=& C_7^{c(3)}(\mu_0=M_W) 
    + \fm{13763}{2187} \ln\fm{\mu_0^2}{M_W^2} 
    + \fm{814}{729}  \ln^2\fm{\mu_0^2}{M_W^2} 
    + {\cal O}(\e),\\[3mm]
C_8^{c(3)}(\mu_0) &=& C_8^{c(3)}(\mu_0=M_W) 
    + \fm{16607}{5832} \ln\fm{\mu_0^2}{M_W^2} 
    + \fm{397}{486} \ln^2\fm{\mu_0^2}{M_W^2} 
    + {\cal O}(\e),\\[3mm]
C_7^{t(1)}(\mu_0) &=&  \left( 1 + \e \ln\fm{\mu_0^2}{m_t^2} 
+ \fm{\e^2}{2} \ln^2\fm{\mu_0^2}{m_t^2} + \fm{\e^2 \pi^2}{12} \right)  
\left( \fm{3 x^3 - 2 x^2}{4 (x - 1)^4} ~\fm{x^\e - 1}{\e} 
     + \fm{22 x^3 - 153 x^2 + 159 x - 46}{72 (x - 1)^3} \right)
\nnb\\ && 
+ \e \left( 1 + \e \ln\fm{\mu_0^2}{m_t^2} \right) 
\left( \fm{-18 x^3 + 150 x^2 - 157 x + 46}{72 (x - 1)^4} ~\fm{x^\e - 1}{\e} 
       + \fm{122 x^3 - 933 x^2 + 975 x - 290}{432 (x - 1)^3} \right)
\nnb\\ && 
+ \e^2 \left( \fm{-108 x^3 + 918 x^2 - 977 x + 290}{432 (x - 1)^4} \ln x
 + \fm{694 x^3 - 5619 x^2 + 5937 x - 1750}{2592 (x - 1)^3} \right) + {\cal O}(\e^3),\\[3mm]
C_8^{t(1)}(\mu_0) &=&  \left( 1 + \e \ln\fm{\mu_0^2}{m_t^2} 
+ \fm{\e^2}{2} \ln^2\fm{\mu_0^2}{m_t^2} + \fm{\e^2 \pi^2}{12} \right)  
\left(   \fm{-3 x^2}{4 (x - 1)^4} ~\fm{x^\e - 1}{\e} 
       + \fm{5 x^3 - 9 x^2 + 30 x - 8}{24 (x - 1)^3} \right)
\nnb\\ && 
+ \e \left( 1 + \e \ln\fm{\mu_0^2}{m_t^2} \right) 
\left(   \fm{-15 x^2 - 14 x + 8}{24 (x - 1)^4} ~\fm{x^\e - 1}{\e} 
       + \fm{13 x^3 + 15 x^2 + 186 x - 88}{144 (x - 1)^3} \right)
\nnb\\ && 
+ \e^2 \left( \fm{-81 x^2 - 130 x + 88}{144 (x - 1)^4}  \ln x
+ \fm{35 x^3 + 273 x^2 + 1110 x - 680}{864 (x - 1)^3} \right) + {\cal O}(\e^3),\\[3mm]
C_7^{t(2)}(\mu_0) &=& \fm{-16 x^4 - 122 x^3 + 80 x^2 - 8 x}{9 (x - 1)^4} 
                       H(x,\e) \left( 1\!+ 2 \e \ln\fm{\mu_0^2}{m_t^2}  \right) 
+\fm{-387 x^4 - 1413 x^3 + 997 x^2 - 65 x + 4}{81 (x - 1)^5} ~\fm{x^\e - 1}{\e} 
\nnb\\ && \hspace{-8mm}  
+\;\fm{94 x^4 + 18665 x^3 - 20682 x^2 + 9113 x - 2006}{486 (x - 1)^4} 
+ \e \left[ \fm{146 x^4 - 4289 x^3 + 2736 x^2 + 14 x - 224}{81 (x - 1)^4} 
            \mbox{Li}_2\!\left(1-\fm{1}{x}\right) 
\right. \nnb\\ && \left. \hspace{-8mm} 
 +\;\fm{-1203 x^4 - 43353 x^3 + 37031 x^2 - 10531 x + 1640}{486 (x - 1)^5}  \ln x 
 +  \fm{-6128 x^4 + 252839 x^3 - 183912 x^2 + 43607 x - 7910}{2916 (x - 1)^4} \right] 
\nnb\\ && \hspace{-8mm} 
+\;\e \ln\fm{\mu_0^2}{m_t^2} \left( 
  \fm{-720 x^4 - 3942 x^3 + 1685 x^2 + 713 x - 220}{81 (x - 1)^5} \ln x
 +\fm{-346 x^4 + 44569 x^3 - 40446 x^2 + 13927 x - 2800}{486 (x - 1)^4} \right)  
\nnb\\ && \hspace{-12mm} 
+\left( \ln\fm{\mu_0^2}{m_t^2}  
\!+\! \fm{3\e}{2} \ln^2\fm{\mu_0^2}{m_t^2} \!+\! \fm{\e\pi^2}{12} \right) 
  \left( \fm{-6 x^4 - 46 x^3 + 28 x^2}{3 (x - 1)^5}  \fm{x^\e - 1}{\e}
     +\fm{34 x^4 + 101 x^3 + 402 x^2 - 397 x + 76}{27 (x - 1)^4} \right) + {\cal O}(\e^2),\\[3mm]
C_8^{t(2)}(\mu_0) &=& \fm{-4 x^4 + 40 x^3 + 41 x^2 + x}{6 (x - 1)^4} 
                      H(x,\e) \left( 1\!+ 2 \e \ln\fm{\mu_0^2}{m_t^2}  \right)  
+ \fm{-144 x^4 + 3177 x^3 + 3661 x^2 + 250 x - 32}{216 (x - 1)^5} ~\fm{x^\e - 1}{\e} 
\nnb\\ && \hspace{-8mm}  
+\;\fm{247 x^4 - 11890 x^3 - 31779 x^2 + 2966 x - 1016}{1296 (x - 1)^4} 
+ \e \left[ \fm{641 x^4 + 184 x^3 + 8001 x^2 - 220 x - 368}{216 (x - 1)^4}  
          \mbox{Li}_2\!\left(1-\fm{1}{x}\right) 
\right. \nnb\\ && \left. \hspace{-8mm} 
 +\;\fm{2982 x^4 + 22581 x^3 + 109751 x^2 - 1018 x - 2968}{1296 (x - 1)^5}  \ln x
 +  \fm{-18557 x^4 - 38590 x^3 - 661839 x^2 - 100078 x + 31096}{7776 (x - 1)^4} \right]
\nnb\\ && \hspace{-8mm} 
+\;\e \ln\fm{\mu_0^2}{m_t^2} \left( 
  \fm{-72 x^4 + 1971 x^3 + 3137 x^2 + 32 x - 100}{54 (x - 1)^5} \ln x
 +\fm{-140 x^4 - 2692 x^3 - 13581 x^2 + 1301 x + 208}{162 (x - 1)^4}  \right)
\nnb\\ && \hspace{-8mm} 
+\left( \ln\fm{\mu_0^2}{m_t^2}  
\!+\! \fm{3\e}{2} \ln^2\fm{\mu_0^2}{m_t^2} \!+\! \fm{\e\pi^2}{12} \right)          
\left( \fm{17 x^3 + 31 x^2}{2 (x - 1)^5} \fm{x^\e - 1}{\e}
    + \fm{35 x^4 - 170 x^3 - 447 x^2 - 338 x + 56}{36 (x - 1)^4} \right) 
    + {\cal O}(\e^2),\\[3mm]
C_7^{t(3)}(\mu_0) &=& C_7^{t(3)}(\mu_0=m_t) 
+ \ln\fm{\mu_0^2}{m_t^2}  
  \left[ \fm{-592 x^5 - 22 x^4 + 12814 x^3 - 6376 x^2 + 512 x}{27 (x - 1)^5} \,
  \mbox{Li}_2\left(1-\fm{1}{x}\right) 
\right. \nnb\\ && \left.
+\fm{-26838 x^5 + 25938 x^4 + 627367 x^3 - 331956 x^2 + 16989 x - 460}{729 (x - 1)^6} \ln x
\right. \nnb\\ && \left.
+ \fm{34400 x^5 + 276644 x^4 - 2668324 x^3 + 1694437 x^2 - 323354 x + 53077}{2187 (x - 1)^5} \right] 
\nnb\\ && 
+ \ln^2\fm{\mu_0^2}{m_t^2} \left[ \fm{-63 x^5 + 532 x^4 + 2089 x^3 - 1118 x^2}{9 (x - 1)^6} \ln x
\right. \nnb\\ && \left.
+ \fm{1186 x^5 - 2705 x^4 - 24791 x^3 - 16099 x^2 + 19229 x - 2740}{162 (x - 1)^5} \right] 
   + {\cal O}(\e),\\[3mm]
C_8^{t(3)}(\mu_0) &=& C_8^{t(3)}(\mu_0=m_t) 
    + \ln\fm{\mu_0^2}{m_t^2} \left[ 
 \fm{-148 x^5 + 1052 x^4 - 4811 x^3 - 3520 x^2 - 61 x}{18 (x - 1)^5} \,
   \mbox{Li}_2\left(1-\fm{1}{x}\right) 
\right. \nnb\\ && \left.
+ \fm{-15984 x^5 + 152379 x^4 - 1358060 x^3 - 1201653 x^2 - 74190 x +  9188}{1944 (x - 1)^6} \ln x
\right. \nnb\\ && \left.
+ \fm{109669 x^5 - 1112675 x^4 + 6239377 x^3 + 8967623 x^2 + 768722 x - 42796}{11664 (x - 1)^5} \right] 
\nnb\\ &&  
    + \ln^2\fm{\mu_0^2}{m_t^2} \left[ 
  \fm{-139 x^4 - 2938 x^3 - 2683 x^2}{12 (x - 1)^6} \ln x 
\right. \nnb\\ && \left.
+ \fm{1295 x^5 - 7009 x^4 + 29495 x^3 + 64513 x^2 + 17458 x - 2072}{216 (x - 1)^5} \right] 
    + {\cal O}(\e).
\eea

As far as the three-loop quantities $C_7^{c(3)}(\mu_0=M_W)$,
$C_8^{c(3)}(\mu_0=M_W)$, $C_7^{t(3)}(\mu_0=m_t)$ and
$C_8^{t(3)}(\mu_0=m_t)$ are concerned, the matching calculation
described in the previous sections gives us expressions for their
expansions at $x\to1$ and $x\to\infty$. Denoting, as before, $z =
1/x$ and $w=1-z$, we find
\bea
C_7^{c(3)}(\mu_0=M_W) &\simeq&  1.525 -0.1165 z +0.01975 z \ln z +0.06283 z^2 +0.005349 z^2 \ln z 
\nnb\\&& 
+0.01005 z^2 \ln^2z -0.04202 z^3 +0.01535 z^3 \ln z -0.00329 z^3 \ln^2 z 
\nnb\\&& 
+0.002372 z^4 -0.0007910 z^4 \ln z +{\cal O}(z^5), \label{nc7c3inf}\\
C_7^{c(3)}(\mu_0=M_W) &\simeq& 1.432 +0.06709 w +0.01257 w^2 +0.004710 w^3 +0.002373w^4 
\nnb\\&& \hspace{-4mm} 
+0.001406w^5 +\!0.0009216 w^6 +\!0.00064730 w^7 +\!0.0004779 w^8 +\!{\cal O}(w^9),\\ 
C_8^{c(3)}(\mu_0=M_W) &\simeq& -1.870 +0.1010 z -0.1218 z \ln z +0.1045 z^2 
-0.03748 z^2 \ln z 
\nnb\\&& 
+0.01151 z^2 \ln^2 z -0.01023 z^3 +0.004342 z^3 \ln z +0.0003031 z^3 \ln^2 z 
\nnb\\&& 
-0.001537 z^4 +0.0007532 z^4 \ln z +{\cal O}(z^5),\\ 
C_8^{c(3)}(\mu_0=M_W) &\simeq& -1.676 -0.1179 w -0.02926 w^2 -0.01297 w^3 -0.007296 w^4
\nnb\\&& 
-0.004672 w^5 -0.003248 w^6 -0.002389 w^7 -0.001831 w^8 +{\cal O}(w^9),\label{nc8c31u}\\
C_7^{t(3)}(\mu_0=m_t) &\simeq& 12.06 +12.93 z +3.013 z \ln z +96.71 z^2 +52.73 z^2 \ln z +147.9 z^3 
\nnb\\&& 
+187.7 z^3 \ln z -144.9 z^4 +236.1 z^4 \ln z +{\cal O}(z^5),\\ 
C_7^{t(3)}(\mu_0=m_t) &\simeq& 11.74 +0.3642 w +0.1155 w^2 -0.003145 w^3 -0.03263 w^4 -0.03528 w^5 
\nnb\\&& 
-0.03076 w^6 -0.02504 w^7 -0.01985 w^8 +{\cal O}(w^9),\\ 
C_8^{t(3)}(\mu_0=m_t) &\simeq& -0.8954 -7.043 z -98.34 z^2 -46.21 z^2 \ln z -127.1 z^3 
\nnb\\&& 
-181.6 z^3 \ln z +535.8 z^4 -76.76 z^4 \ln z +{\cal O}(z^5),\\ 
C_8^{t(3)}(\mu_0=m_t) &\simeq& -0.6141 -0.8975 w -0.03492 w^2 +0.06791 w^3 +0.07966 w^4 
\nnb\\&& 
+0.07226 w^5 +0.06132 w^6 +0.05096 w^7 +0.04216 w^8 +{\cal O}(w^9).
\eea
\mathindent1cm
While only numerical values of the expansion coefficients have been
given above, their exact values can easily be found from similar
expansions for the unrenormalized three-loop results (Appendix~A) and
from the formulae of
Sections~\ref{sec:eff.theory}--\ref{sec:matching}.
\begin{figure}[t]
\vspace*{-3cm}
\begin{center}
\hspace*{-4cm}
\includegraphics[width=9cm,angle=0]{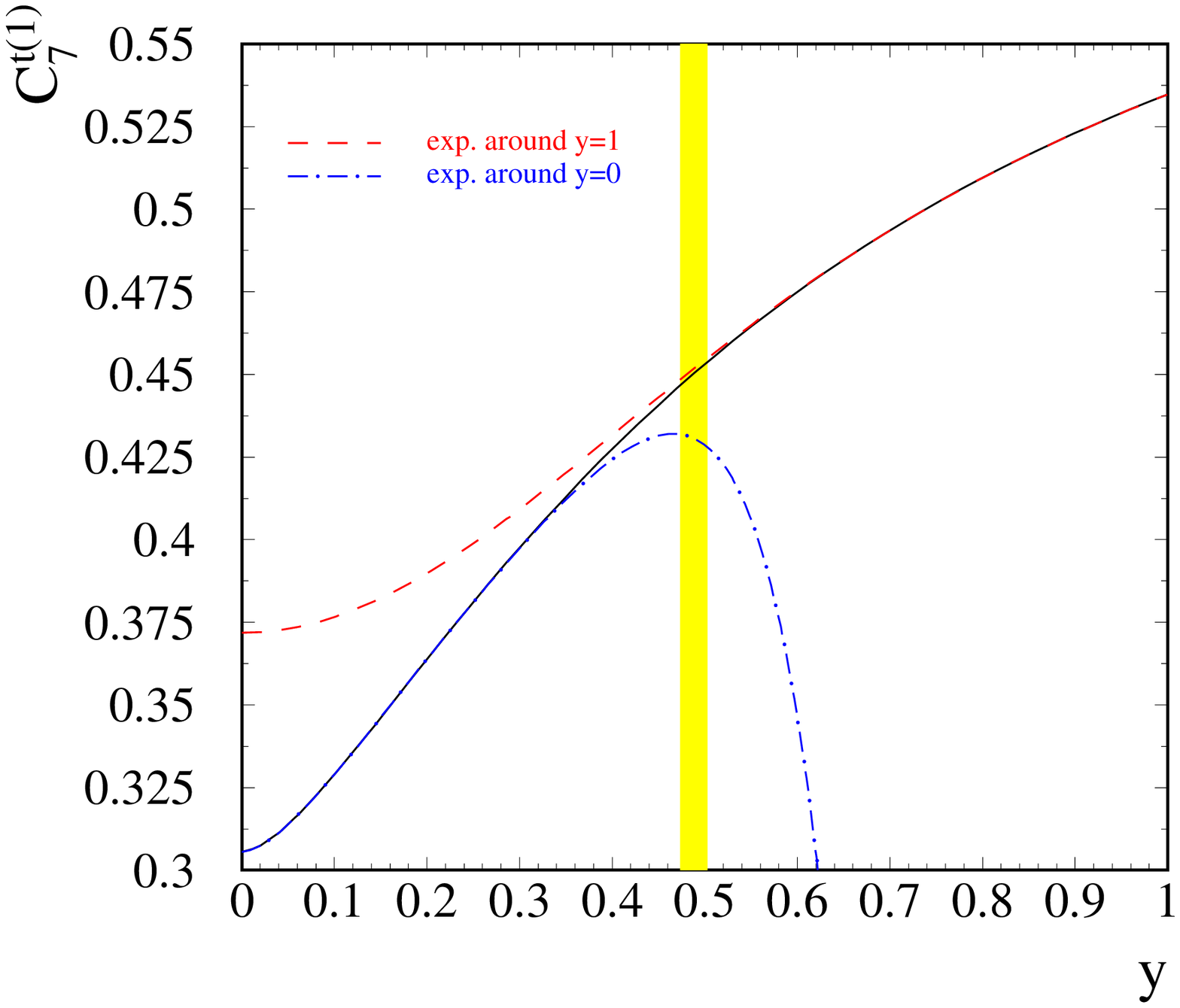}\hspace{-7mm}
\includegraphics[width=9cm,angle=0]{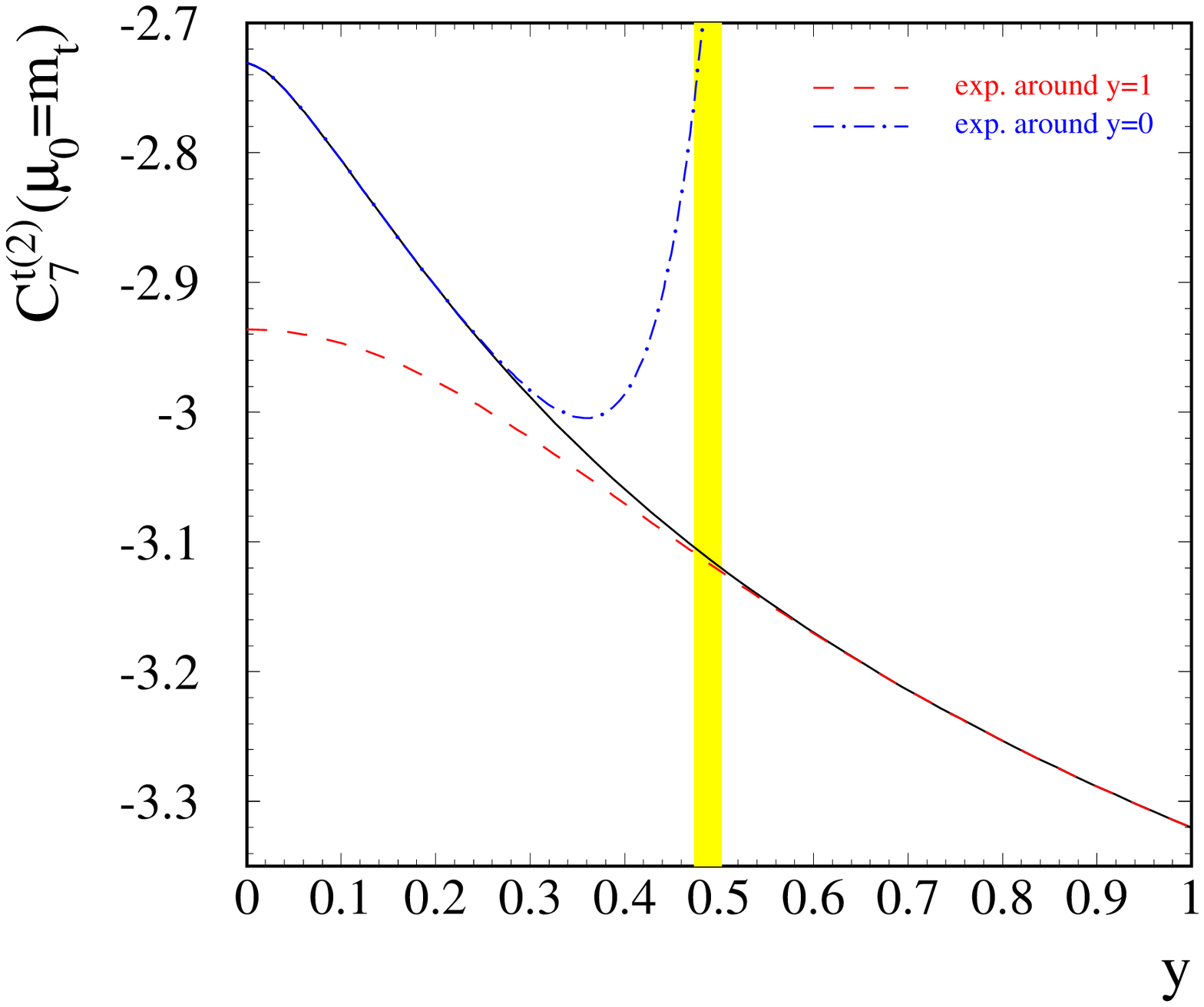}\hspace{-4cm} \ \\
\hspace*{-4cm}
\includegraphics[width=9cm,angle=0]{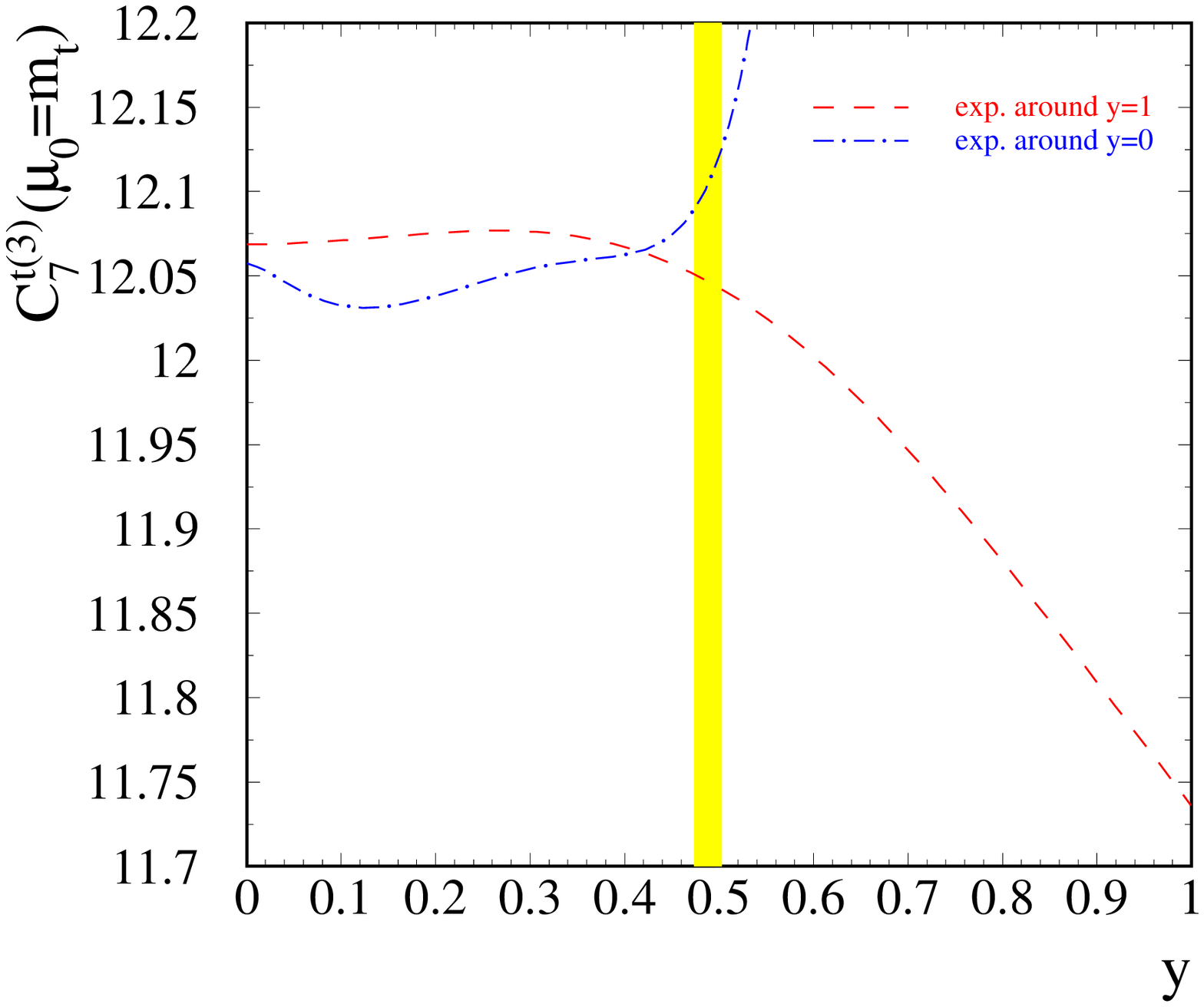}\hspace{-7mm}
\includegraphics[width=9cm,angle=0]{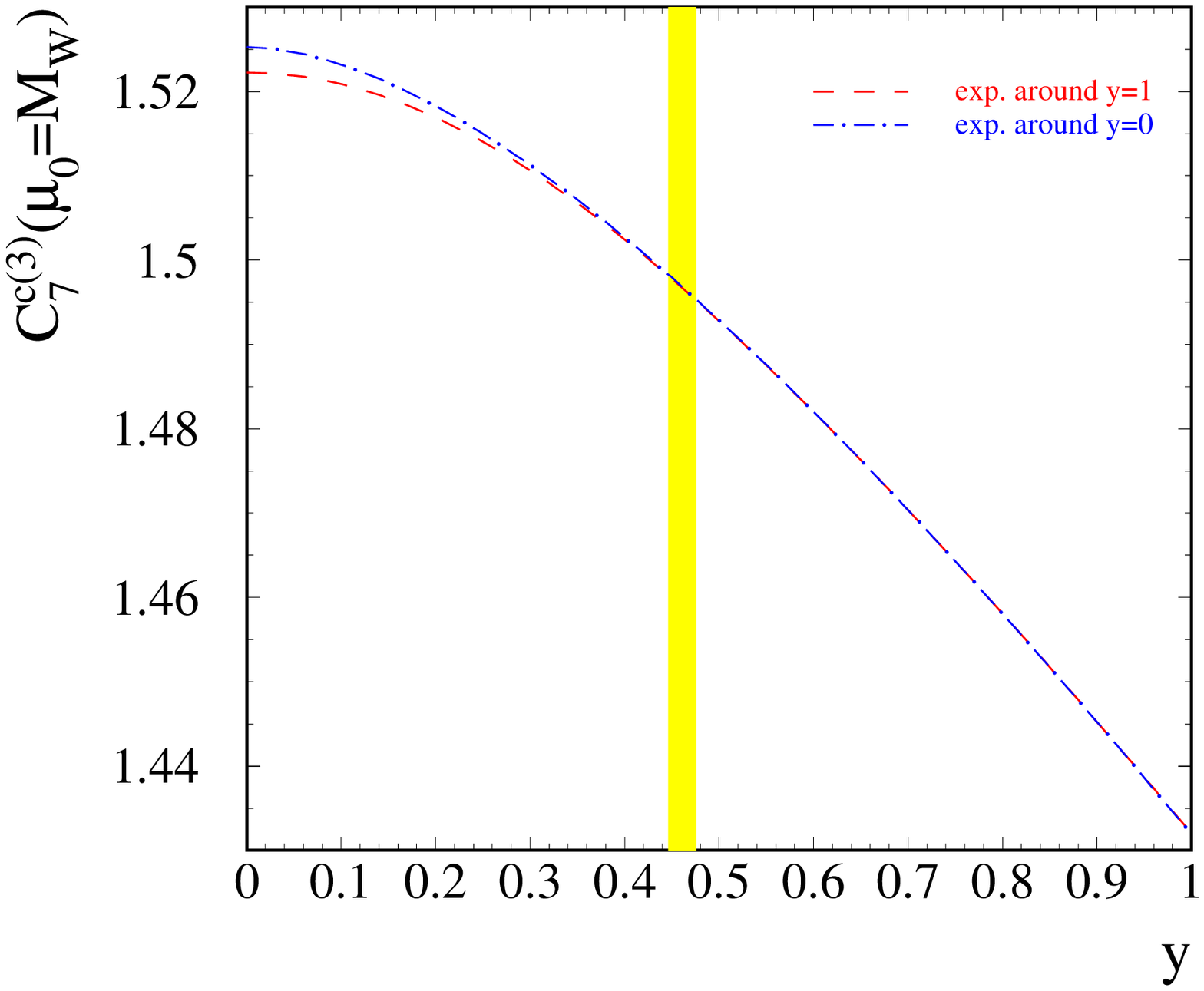}\hspace{-4cm} \ 
\caption{\sf The coefficients $C_7^{Q(n)}(\mu_0)$ as functions of $y =
M_W/m_t(\mu_0)$. The (blue) dot-dashed lines correspond to their
expansions in $y$ up to $y^8$. The (red) dashed lines describe the
expansions in $(1-y^2)$ up to $(1-y^2)^8$. The (black) solid lines in
the one- and two-loop cases correspond to the known exact
expressions. The (yellow) vertical strips indicate the experimental
range for $y$.} 
\label{fig:c7}
\end{center}
\vspace*{-5mm}
\end{figure}

\begin{figure}[t]
\vspace*{-3cm}
\begin{center}
\hspace*{-4cm}
\includegraphics[width=9cm,angle=0]{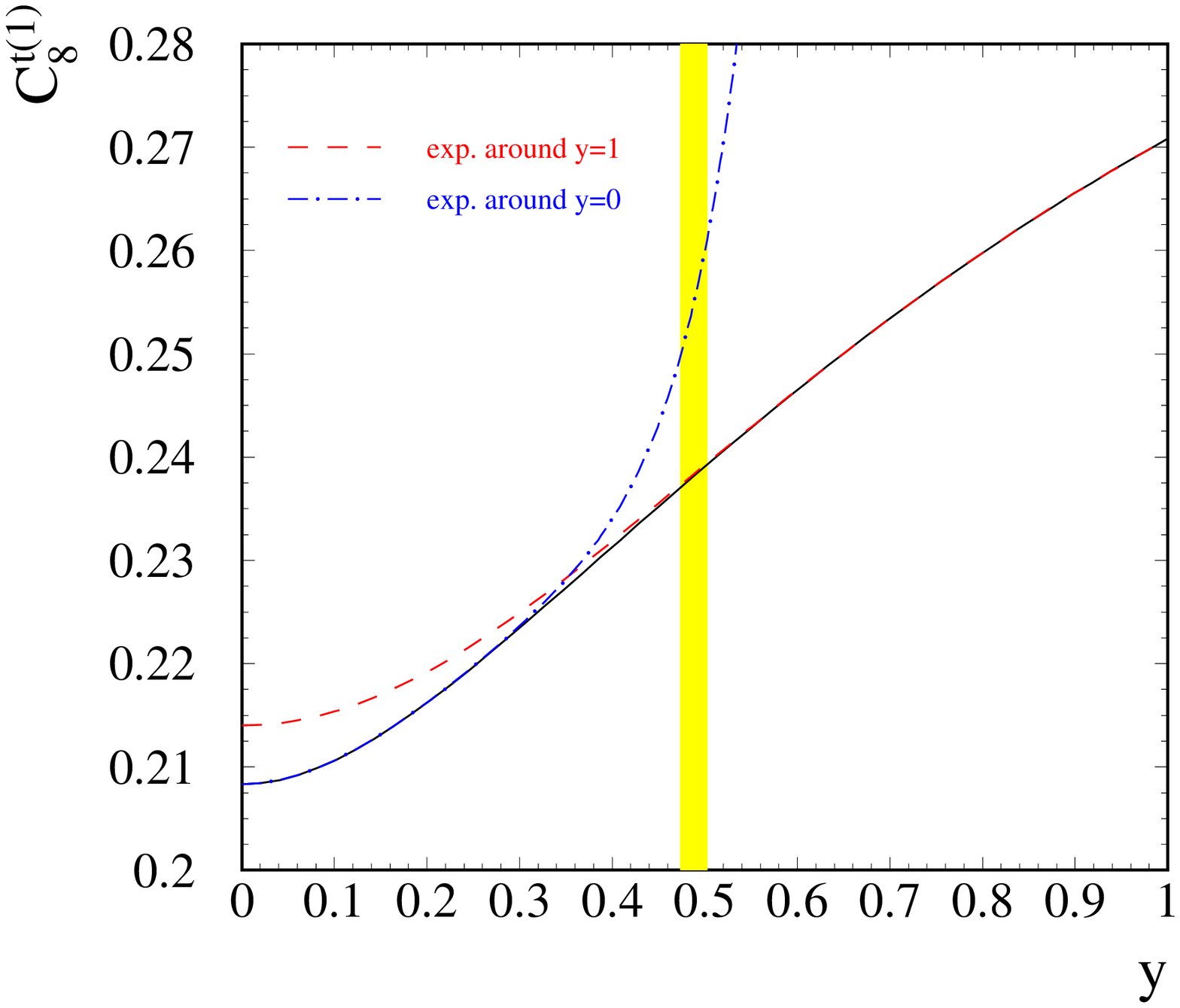}\hspace{-7mm}
\includegraphics[width=9cm,angle=0]{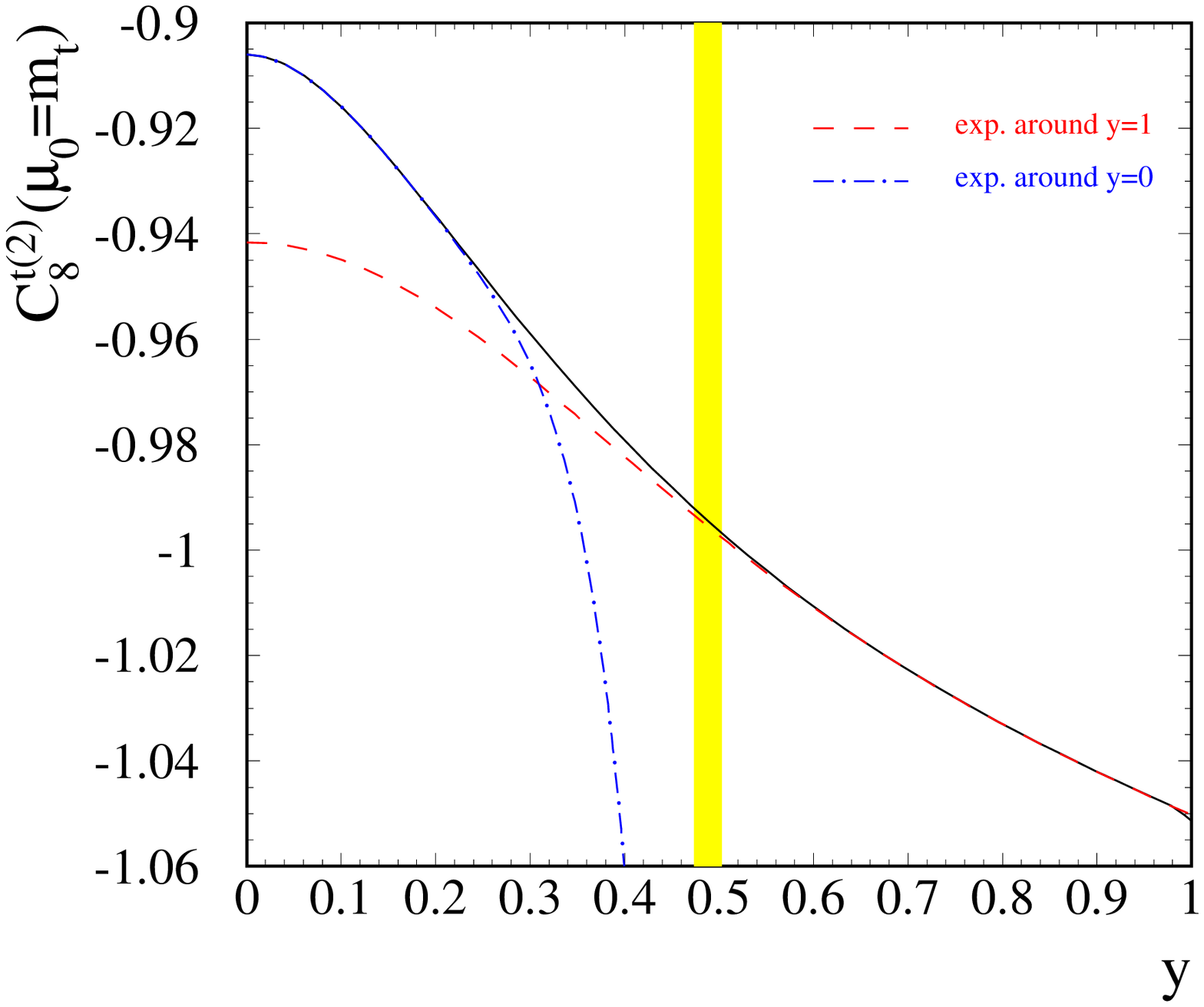}\hspace{-4cm} \ \\
\hspace*{-4cm}
\includegraphics[width=9cm,angle=0]{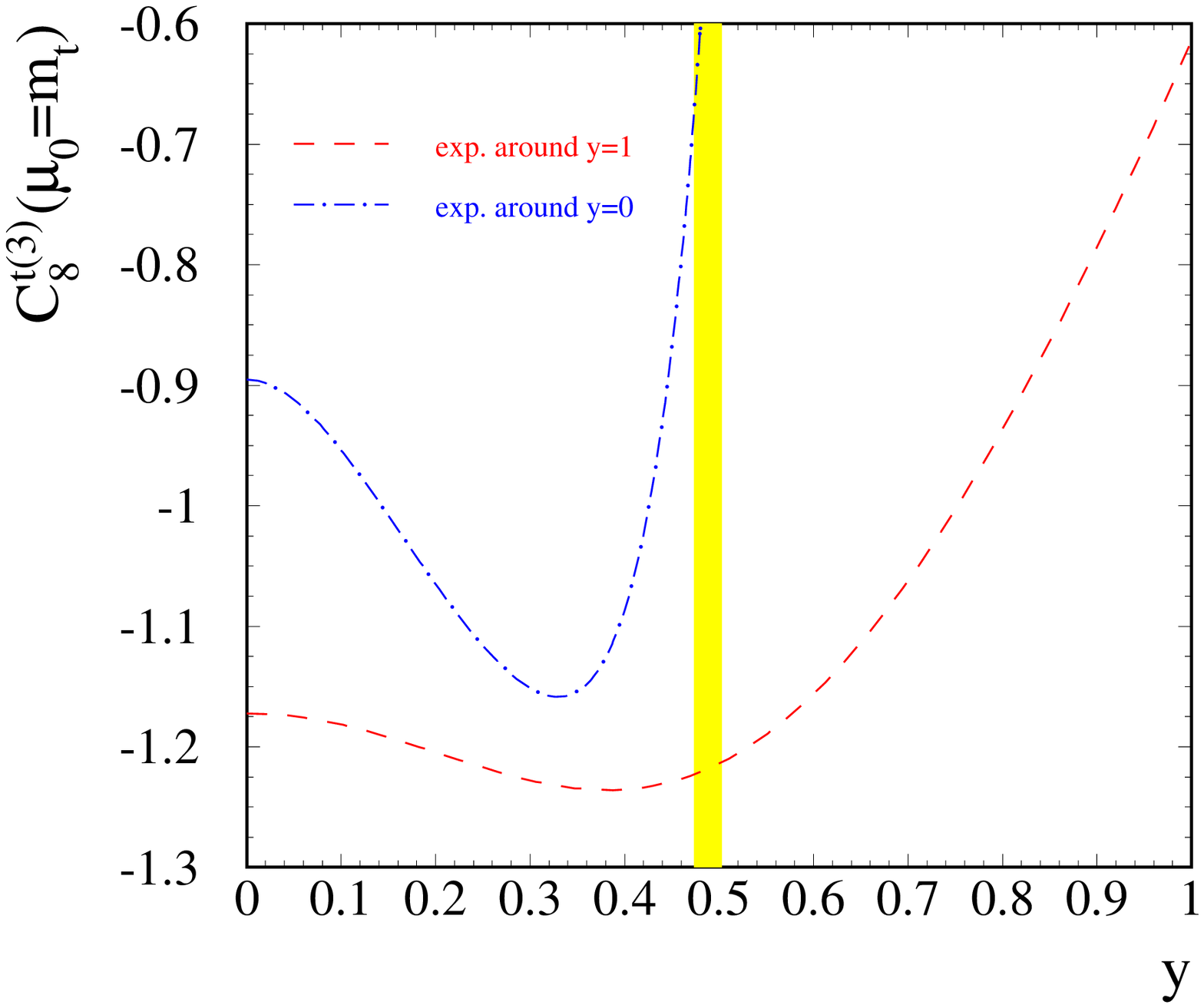}\hspace{-7mm}
\includegraphics[width=9cm,angle=0]{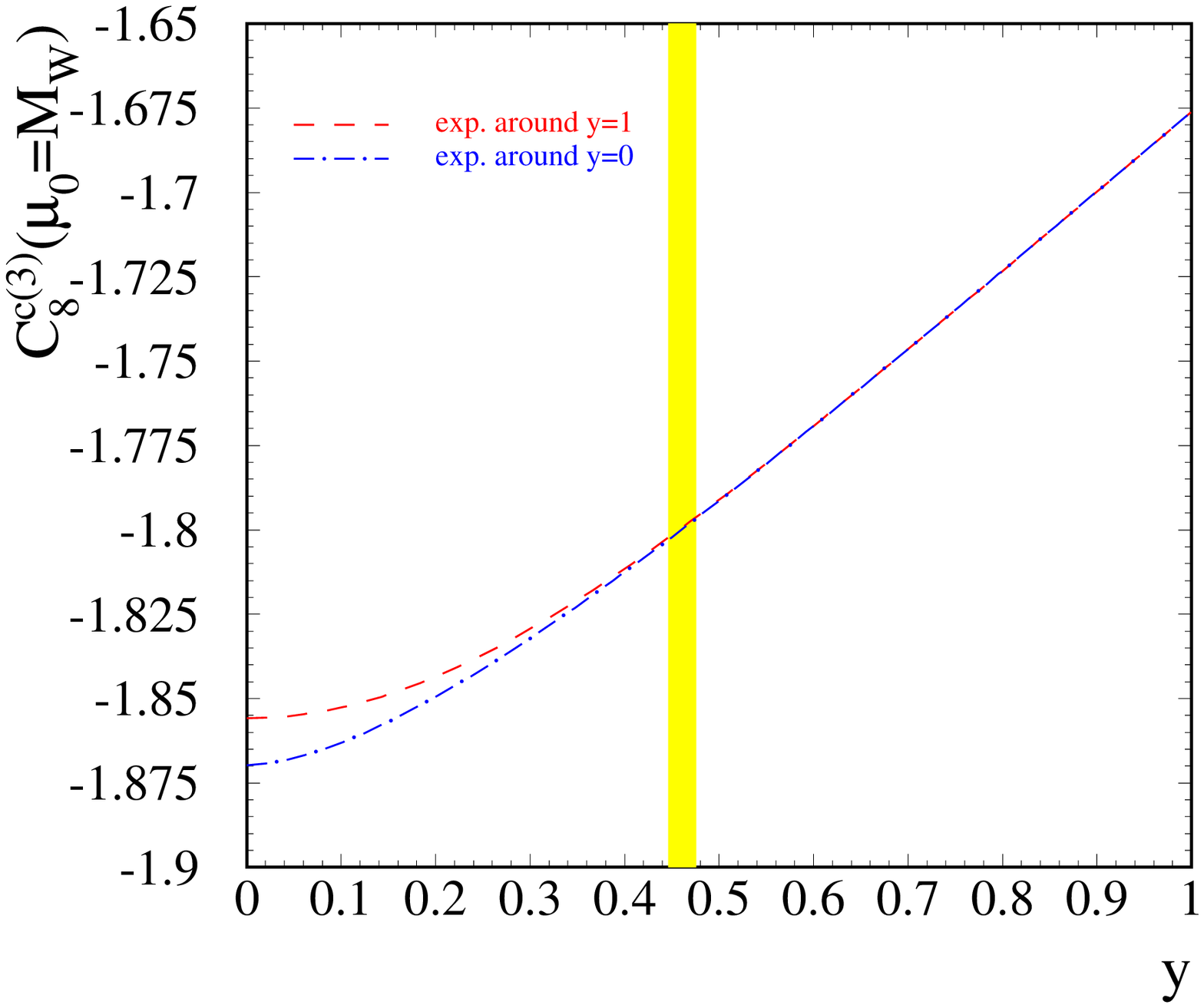}\hspace{-4cm} \ 
\caption{\sf Same as Fig.~\ref{fig:c7} but for $C_8^{Q(n)}(\mu_0)$.} 
\label{fig:c8}
\end{center}
\vspace*{-5mm}
\end{figure}

In Figs.~\ref{fig:c7} and \ref{fig:c8}, the top-mass dependent
coefficients $C_i^{t(n)}(\mu_0=m_t)$ and $C_i^{c(3)}(\mu_0=M_W)$ for
$i=7,8$ are plotted as functions of $y = M_W/m_t(\mu_0)$. The
different choice of renormalization scales in the top and charm
sectors allows us to avoid logarithmic divergences at large $m_t$ and,
consequently, achieve better control over the behaviour of the
expansions. This is the main reason why $\mu_0$ has been normalized to
$M_W$ in the charm sector and to $m_t$ in the top sector, in all our
intermediate and final expressions.\footnote{
Apart from that, many of the top-sector expressions would be
significantly longer if $\mu_0$ was normalized to $M_W$ there.}

The variable $y$ changes from 0 to 1, i.e. both starting points of our
expansions are present in the figures.  Note the relatively narrow
ranges of the coefficient values on the vertical axes. The large $m_t$
expansions (up to $y^8$) are depicted by the dot-dashed lines, while
the expansions around $m_t = M_W$ (up to $(1-y^2)^8$) are given by the
dashed ones.  In the one- and two-loop cases, solid curves show the
exact results.  The vertical strips mark the experimental values for
$y$ that we take $(0.488 \pm 0.015)$ for $\mu_0=m_t$, and $(0.461 \pm
0.015)$ for $\mu_0=M_W$.

Comparing the three curves in the one- and two-loop cases (the two
upper plots in both figures), one can conclude that a combination of
the two expansions at hand gives a good determination of the studied
coefficients in the whole considered range of $y$. However, the
expansion starting from $y=1$ works somewhat better for the physical
values of $m_t$ and $M_W$.  Most probably, including more terms in the
the large $m_t$ expansion could improve its behaviour around $y=0.5$.

Although we do not know the exact curves in the three-loop case, the
same pattern seems to repeat. In fact, the charm-sector expansions
perfectly overlap in the physical region. In the top sector, one can
(conservatively) conclude that
\bea
C_7^{t(3)}(\mu_0=m_t) &=& 12.05 \pm 0.05,\\
C_8^{t(3)}(\mu_0=m_t) &=& -1.2 \pm 0.1,
\eea 
which is perfectly accurate for any phenomenological application. Let
us note that a change of $C_7^{t(3)}(\mu_0=m_t)$ from 12 to 13 would
affect the $b \to s \gamma$ decay width by only 0.02\%, while a
similar variation of $C_8^{t(3)}(\mu_0=m_t)$ would cause even a smaller
effect.

For the three-loop charm-sector coefficients, the uncertainty from the
expansions is smaller than the one from the experimental error in
$m_t$. Thus, one can safely use Eqs.~(\ref{nc7c3inf})--(\ref{nc8c31u})
as they stand, without any additional uncertainty. Accurate values in
the range $0.4 < y < 0.6$ can also be found from the following fits:
\bea
C_7^{c(3)}(\mu_0=M_W) &=&  1.458 \left( \f{m_t}{M_W} \right)^{0.0338},\\
C_8^{c(3)}(\mu_0=M_W) &=& -1.718 \left( \f{m_t}{M_W} \right)^{0.0598}.
\eea 

\begin{figure}[h]
\vspace*{-5mm}
\begin{center}
\hspace*{-4cm}
\includegraphics[width=9cm,angle=0]{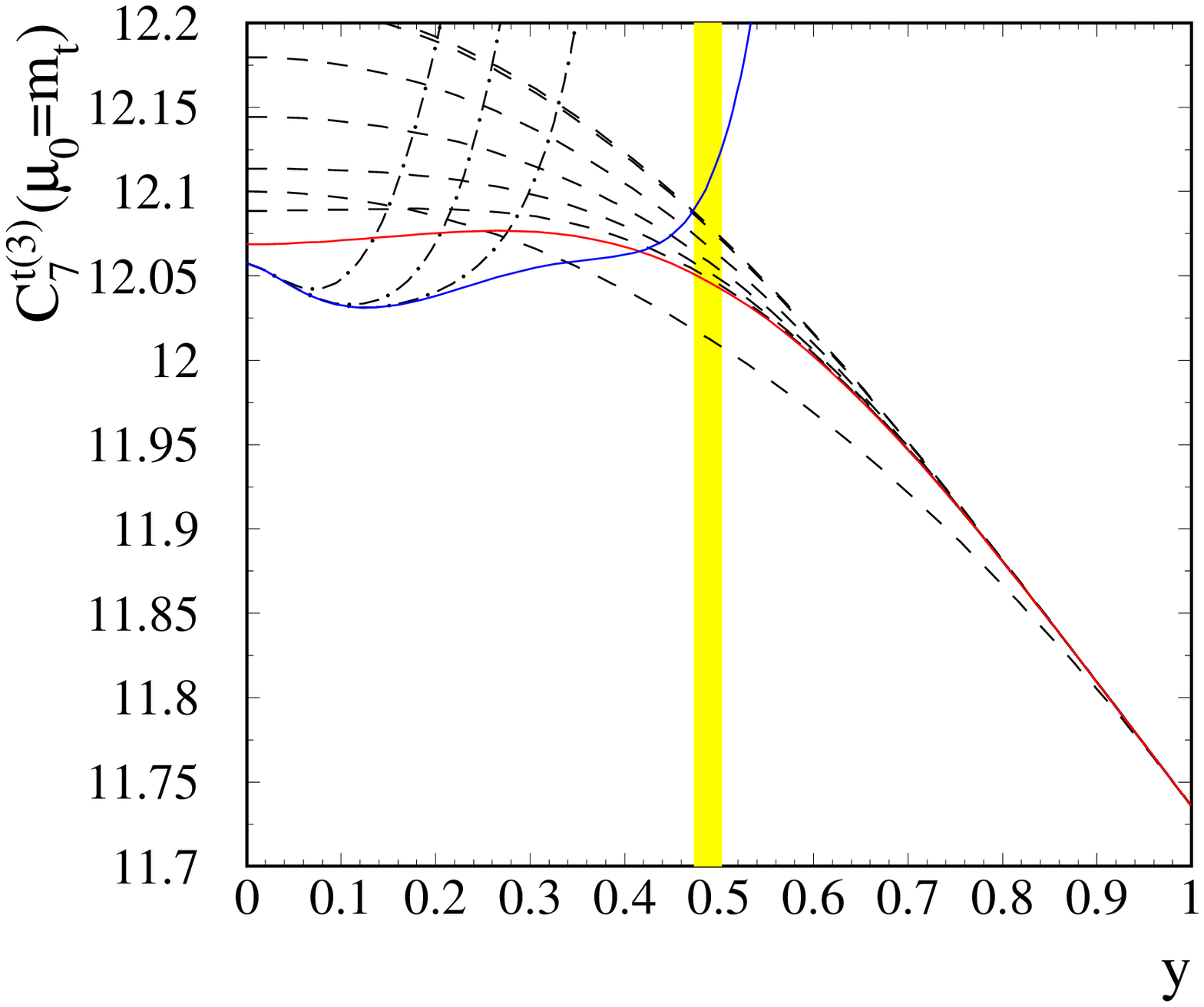}\hspace{-7mm}
\includegraphics[width=9cm,angle=0]{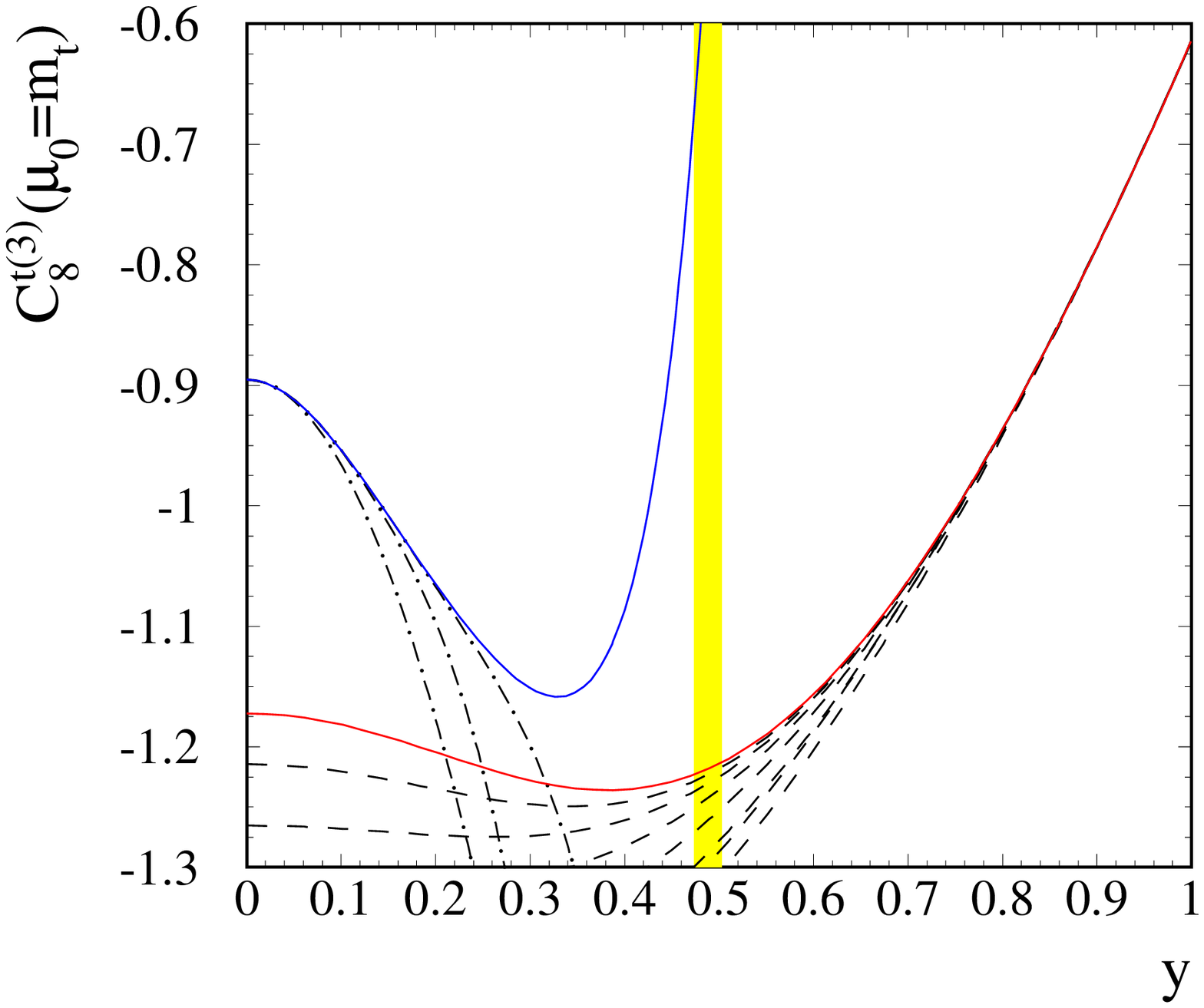}\hspace{-4cm} \ 
\caption{\sf The three-loop top-sector coefficients.  The solid lines
represent the highest orders we know (as in Figs.~\ref{fig:c7} and
\ref{fig:c8}). The dashed and dot-dashed lines show the lower orders.}
\label{fig:ct.low}
\end{center}
\vspace*{-5mm}
\end{figure}

It is instructive to study the behaviour of the three-loop top-sector
coefficients in a plot where subsequent terms of our expansions are
successively taken into account. This is shown in Fig.~\ref{fig:ct.low}.
The quality of the two expansions in various regions of $y$ is
transparent there.

\newpage

\newsection{Conclusions \label{sec:conclusions}}

The three-loop matching conditions found in the present paper complete
the first out of three steps (matching, mixing and matrix elements)
that are necessary for finding the NNLO QCD corrections to $\bar{B}
\to X_s \gamma$.  The effect of the NNLO matching alone is scheme- and
scale-dependent. In the $\overline{\rm MS}$ scheme with $M_W < \mu_0 <
m_t$, it stays within 2\% of the decay width, i.e. it is significantly
smaller than the total higher-order perturbative uncertainty that was
estimated in Ref.~\cite{Gambino:2001ew}. This uncertainty is expected
to get significantly suppressed in the near future, after the
remaining two steps of the NNLO calculation are performed.

The methods that we have applied in the present work are, in
principle, applicable to any three-loop matching computation involving
several different mass scales. A detailed description that we have
presented for each step of our procedure can serve as a guideline for
treating similar problems in various domains of particle
phenomenology.

\section*{Acknowledgements}

M.M. is grateful to Ben Moore, Joachim Stadel and Daniel Wyler for
helpful discussions and advice concerning the Z-Box computer at the
University of Z\"urich.  He acknowledges support from the
Schweizerischer Nationalfonds, from the Polish Committee for
Scientific Research under the grant 2~P03B~121~20, and from the
European Community's Human Potential Programme under the contract
HPRN-CT-2002-00311, EURIDICE.

\newappendix{Appendix A:~~ Three-loop expansion terms}
\def\theequation{A.\arabic{equation}}

In this appendix, we present our results for the coefficients
$a_{nm}^{kQ}$ and $b_{n}^{kQ}$ from Eqs.~(\ref{exp.f.a}) and~(\ref{exp.f.b})
up to $n=4$ and $n=8$, respectively.  They are given in terms of the
following symbols (see also Eq. (16) of
Ref.~\cite{Steinhauser:2000ry}):
\begin{eqnarray}
  \Dthree &=&
  6\zeta_3
  -\frac{15}{4}\zeta_4 
  -6 \left[\mbox{Cl}_2\left(\frac{\pi}{3}\right) \right]^2 
  \,,
  \nonumber\\
  \Bfour &=& 
  - 4 \zeta_2 \ln^2 2
  + \frac{2}{3}\ln^4 2 
  - \frac{13}{2} \zeta_4
  + 16 \mbox{Li}_4\left(\frac{1}{2}\right) 
  \,,
  \nonumber\\
  \Stwo &=& \frac{4}{9\sqrt{3}} \mbox{Cl}_2\left(\frac{\pi}{3}\right)
  \,,
  \nonumber\\
  \OepStwo &=&
  - \frac{763}{32}
  - \frac{9\pi\sqrt{3}\ln^2 3}{16}
  - \frac{35\pi^3\sqrt{3}}{48}
  + \frac{195}{16}\zeta_2
  - \frac{15}{4}\zeta_3
  + \frac{57}{16}\zeta_4
  \nonumber\\&&\mbox{}
  + \frac{45\sqrt{3}}{2} {\rm Cl}_2\left(\frac{\pi}{3}\right)
  - 27\sqrt{3} {\rm Im}\left[ {\rm
     Li}_3\left(\frac{e^{-i\pi/6}}{\sqrt{3}}\right)\right]
  \,,
  \nonumber\\
  \Toneep &=&
  - \frac{45}{2} 
  - \frac{\pi\sqrt{3} \ln^2 3}{8} 
  \nonumber\\&&\mbox{}
  - \frac{35\pi^3\sqrt{3}}{216}
  - \frac{9}{2}\zeta_2
  + \zeta_3
  + 6\sqrt{3}{\rm Cl}_2\left(\frac{\pi}{3}\right)
  - 6\sqrt{3} {\rm Im}\left[ {\rm
      Li}_3\left(\frac{e^{-i\pi/6}}{\sqrt{3}}\right)\right] 
  \label{eq:master}
  \,,
\end{eqnarray}
where ${\rm Cl}_2(x) = {\rm Im}\left[ {\rm Li}_2 \left(e^{ix} \right) \right]$.\\

The expansion coefficients that we have found read
\mathindent0cm
\begin{eqnarray}
  a_{00}^{7t} &=&
    +\fracown{70\zetathree}{243}
    +\fracown{1587\Stwo}{280}
    +\fracown{43\pi^4}{405}
    +\fracown{416341\pi^2}{612360}
    +\fracown{46\Dthree}{81}
    -\fracown{92\Bfour}{81}
    +\fracown{820640533}{9185400}
    \,,
  \nonumber\\
  a_{10}^{7t} &=&
    +\fracown{307721\zetathree}{324}
    -\fracown{67\Toneep}{18}
    -\fracown{284327\Stwo}{840}
    +\fracown{96959\pi^4}{116640}
    -\fracown{53880251\pi^2}{816480}
    +\fracown{67\OepStwo}{27}
    +\fracown{680\Dthree}{81}
    -\fracown{1360\Bfour}{81}
    -\fracown{2469729799}{4082400}
    \,,
  \nonumber\\
  a_{11}^{7t} &=&
    -\fracown{201\Stwo}{4}
    +\fracown{49\pi^2}{54}
    -\fracown{2669}{810}
    \,,
  \nonumber\\
  a_{12}^{7t} &=&
    -\fracown{11}{6}
    \,,
  \nonumber\\
  a_{13}^{7t} &=&
    -\fracown{19}{18}
    \,,
  \nonumber\\
  a_{20}^{7t} &=&
    +\fracown{138245\zetathree}{54}
    -\fracown{4073\Toneep}{81}
    -\fracown{333063\Stwo}{140}
    +\fracown{122821\pi^4}{58320}
    -\fracown{7316857\pi^2}{81648}
    +\fracown{8146\OepStwo}{243}
    + 34\Dthree
    -68\Bfour
    -\fracown{1981904129}{408240}
    \,, \hspace{-3cm}
  \nonumber\\
  a_{21}^{7t} &=&
    -\fracown{4073\Stwo}{6}
    +\fracown{3943\pi^2}{162}
    +\fracown{306769}{1215}
    \,,
  \nonumber\\
  a_{22}^{7t} &=&
    +\fracown{4613}{81}
    \,,
  \nonumber\\
  a_{23}^{7t} &=&
    +\fracown{146}{27}
    \,,
  \nonumber\\
  a_{30}^{7t} &=&
    +\fracown{138120863\zetathree}{26244}
    -\fracown{3547685\Toneep}{13122}
    -\fracown{98842253\Stwo}{9720}
    +\fracown{17476801\pi^4}{9447840}
    +\fracown{284448283\pi^2}{3149280}
    +\fracown{3547685\OepStwo}{19683}
\nnb\\ && \hspace{9cm}
    +\fracown{7432\Dthree}{81}
    -\fracown{14864\Bfour}{81}
    -\fracown{32458492807}{1574640}
    \,,
  \nonumber\\
  a_{31}^{7t} &=&
    -\fracown{3547685\Stwo}{972}
    +\fracown{506753\pi^2}{2916}
    +\fracown{46342189}{17496}
    \,,
  \nonumber\\
  a_{32}^{7t} &=&
    +\fracown{1751809}{2916}
    \,,
  \nonumber\\
  a_{33}^{7t} &=&
    +\fracown{251}{3}
    \,,
  \nonumber\\
  a_{40}^{7t} &=&
    +\fracown{257322953\zetathree}{26244}
    -\fracown{12491099\Toneep}{13122}
    -\fracown{5628051553\Stwo}{174960}
    -\fracown{54918881\pi^4}{9447840}
    +\fracown{12685755337\pi^2}{22044960}
    +\fracown{12491099\OepStwo}{19683}
\nnb\\ && \hspace{9cm}
    +\fracown{16166\Dthree}{81}
    -\fracown{32332\Bfour}{81}
    -\fracown{190409709691}{3149280}
    \,,
  \nonumber\\
  a_{41}^{7t} &=&
    -\fracown{12491099\Stwo}{972}
    +\fracown{1978619\pi^2}{2916}
    +\fracown{531316991}{43740}
    \,,
  \nonumber\\
  a_{42}^{7t} &=&
    +\fracown{664799}{243}
    \,,
  \nonumber\\
  a_{43}^{7t} &=&
    +\fracown{3424}{9}
    \,,
\end{eqnarray}
\begin{eqnarray}
  b_{0}^{7t} &=&
    -\fracown{2901893\zetathree}{38880}
    +\fracown{4909\pi^2}{2160}
    +\fracown{1797615371}{8748000}
    \,,
  \nonumber\\
  b_{1}^{7t} &=&
    +\fracown{94143997\zetathree}{1244160}
    -\fracown{6763\pi^2}{15120}
    -\fracown{113487750073}{979776000}
    \,,
  \nonumber\\
  b_{2}^{7t} &=&
    +\fracown{91942073\zetathree}{860160}
    -\fracown{241\pi^2}{1512}
    -\fracown{10092619036343}{76814438400}
    \,,
  \nonumber\\
  b_{3}^{7t} &=&
    +\fracown{137418234607\zetathree}{891813888}
    -\fracown{8\pi^2}{135}
    -\fracown{13709395882765691}{73741860864000}
    \,,
  \nonumber\\
  b_{4}^{7t} &=&
    +\fracown{7490373009073\zetathree}{35672555520}
    -\fracown{593\pi^2}{30240}
    -\fracown{5960644239577417}{23597395476480}
    \,,
  \nonumber\\
  b_{5}^{7t} &=&
    +\fracown{467301421361\zetathree}{1698693120}
    -\fracown{131\pi^2}{47520}
    -\fracown{6438103242654889429}{19467851268096000}
    \,,
  \nonumber\\
  b_{6}^{7t} &=&
    +\fracown{329068267226885\zetathree}{941755465728}
    +\fracown{1139\pi^2}{249480}
    -\fracown{1798715398515135307759}{4282927278981120000}
    \,,
  \nonumber\\
  b_{7}^{7t} &=&
    +\fracown{1164930029277053\zetathree}{2690729902080}
    +\fracown{353\pi^2}{46332}
    -\fracown{445971686554467633047}{857084922534297600}
    \,,
  \nonumber\\
  b_{8}^{7t} &=&
    +\fracown{32688338029429333\zetathree}{62185757736960}
    +\fracown{56293\pi^2}{6486480}
    -\fracown{12555120069922446322011879253}{19873056681818251591680000}
    \,,
\end{eqnarray}
\begin{eqnarray}
  a_{00}^{8t} &=&
    -\fracown{22301\zetathree}{648}
    +\fracown{22149\Stwo}{224}
    +\fracown{17\pi^4}{2160}
    +\fracown{170659\pi^2}{25515}
    +\fracown{13\Dthree}{216}
    -\fracown{13\Bfour}{108}
    +\fracown{1189623529}{61236000}
    \,,
  \nonumber\\
  a_{10}^{8t} &=&
    -\fracown{147193\zetathree}{432}
    +\fracown{1075273\Stwo}{2240}
    -\fracown{1733\pi^4}{2430}
    +\fracown{3172381\pi^2}{544320}
    -\fracown{56\Dthree}{27}
    +\fracown{112\Bfour}{27}
    +\fracown{409119733}{1360800}
    \,,
  \nonumber\\
  a_{11}^{8t} &=&
    -\fracown{581}{5400}
    \,,
  \nonumber\\
  a_{12}^{8t} &=&
    -\fracown{1}{80}
    \,,
  \nonumber\\
  a_{20}^{8t} &=&
    -\fracown{304559\zetathree}{216}
    +\fracown{49\Toneep}{2}
    +\fracown{3422759\Stwo}{1920}
    -\fracown{9289\pi^4}{4320}
    +\fracown{271907\pi^2}{31104}
    -\fracown{49\OepStwo}{3}
    -\fracown{349\Dthree}{24}
    +\fracown{349\Bfour}{12}
    +\fracown{487461187}{155520}
    \,,
  \nonumber\\
  a_{21}^{8t} &=&
    +\fracown{1323\Stwo}{4}
    -\fracown{3811\pi^2}{432}
    -\fracown{8187697}{64800}
    \,,
  \nonumber\\
  a_{22}^{8t} &=&
    -\fracown{63517}{2160}
    \,,
  \nonumber\\
  a_{23}^{8t} &=&
    -\fracown{199}{72}
    \,,
  \nonumber\\
  a_{30}^{8t} &=&
    -\fracown{8128418\zetathree}{2187}
    +\fracown{3891425\Toneep}{17496}
    +\fracown{284508347\Stwo}{36288}
    -\fracown{26167261\pi^4}{12597120}
    +\fracown{115972585\pi^2}{5878656}
    -\fracown{3891425\OepStwo}{26244}
\nnb\\ && \hspace{9cm}
    -\fracown{6097\Dthree}{108}
    +\fracown{6097\Bfour}{54}
    +\fracown{255386869021}{14696640}
    \,,
  \nonumber\\
  a_{31}^{8t} &=&
    +\fracown{3891425\Stwo}{1296}
    -\fracown{442091\pi^2}{3888}
    -\fracown{1407902803}{583200}
    \,,
  \nonumber\\
  a_{32}^{8t} &=&
    -\fracown{10526363}{19440}
    \,,
  \nonumber\\
  a_{33}^{8t} &=&
    -\fracown{117}{2}
    \,,
  \nonumber\\
  a_{40}^{8t} &=&
    -\fracown{1201430399\zetathree}{139968}
    +\fracown{72196517\Toneep}{69984}
    +\fracown{28165051597\Stwo}{933120}
    +\fracown{83544979\pi^4}{10077696}
    +\fracown{33972092933\pi^2}{117573120}
    -\fracown{72196517\OepStwo}{104976}
\nnb\\ && \hspace{9cm}
    -\fracown{34039\Dthree}{216}
    +\fracown{34039\Bfour}{108}
    +\fracown{206714107565}{3359232}
    \,,
  \nonumber\\
  a_{41}^{8t} &=&
    +\fracown{72196517\Stwo}{5184}
    -\fracown{9280985\pi^2}{15552}
    -\fracown{4347341779}{291600}
    \,,
  \nonumber\\
  a_{42}^{8t} &=&
    -\fracown{20676181}{6480}
    \,,
  \nonumber\\
  a_{43}^{8t} &=&
    -\fracown{1379}{4}
    \,,
\end{eqnarray}
\begin{eqnarray}
  b_{0}^{8t} &=&
    -\fracown{3426427\zetathree}{207360}
    +\fracown{4007\pi^2}{2880}
    +\fracown{383324521}{4665600}
    \,,
  \nonumber\\
  b_{1}^{8t} &=&
    +\fracown{377401\zetathree}{12960}
    -\fracown{901\pi^2}{10080}
    -\fracown{683934529}{16329600}
    \,,
  \nonumber\\
  b_{2}^{8t} &=&
    +\fracown{257020361\zetathree}{3870720}
    +\fracown{247\pi^2}{20160}
    -\fracown{5133539931187}{64012032000}
    \,,
  \nonumber\\
  b_{3}^{8t} &=&
    +\fracown{148678249549\zetathree}{1486356480}
    +\fracown{25\pi^2}{1008}
    -\fracown{590349164605337}{4916124057600}
    \,,
  \nonumber\\
  b_{4}^{8t} &=&
    +\fracown{48644809387\zetathree}{339738624}
    +\fracown{869\pi^2}{40320}
    -\fracown{33797460512670161}{196644962304000}
    \,,
  \nonumber\\
  b_{5}^{8t} &=&
    +\fracown{3067923823757\zetathree}{15854469120}
    +\fracown{3649\pi^2}{221760}
    -\fracown{6031703724292608407}{25957135024128000}
    \,,
  \nonumber\\
  b_{6}^{8t} &=&
    +\fracown{787852603809259\zetathree}{3139184885760}
    +\fracown{16189\pi^2}{1330560}
    -\fracown{9469141155025123085807}{31408133379194880000}
    \,,
  \nonumber\\
  b_{7}^{8t} &=&
    +\fracown{113217484992547\zetathree}{358763986944}
    +\fracown{173\pi^2}{19305}
    -\fracown{1857857271774679586364179}{4899668807154401280000}
    \,,
  \nonumber\\
  b_{8}^{8t} &=&
    +\fracown{96339777793582171\zetathree}{248743030947840}
    +\fracown{11485\pi^2}{1729728}
    -\fracown{948680668305509273934263213}{2038262223776230932480000}
    \,,
\end{eqnarray}
\begin{eqnarray}
  a_{00}^{7c} &=&
    -\fracown{5032\zetathree}{243}
    +\fracown{29683\pi^2}{4374}
    -\fracown{517861}{39366}
    \,,
  \nonumber\\
  a_{01}^{7c} &=&
    +\fracown{407}{729}
    \,,
  \nonumber\\
  a_{02}^{7c} &=&
    -\fracown{112}{243}
    \,,
  \nonumber\\
  a_{10}^{7c} &=&
    -\fracown{112}{1215}
    \,,
  \nonumber\\
  a_{11}^{7c} &=&
    +\fracown{8}{405}
    \,,
  \nonumber\\
  a_{20}^{7c} &=&
    +\fracown{19\pi^2}{5670}
    +\fracown{4960261}{166698000}
    \,,
  \nonumber\\
  a_{21}^{7c} &=&
    -\fracown{2123}{396900}
    \,,
  \nonumber\\
  a_{22}^{7c} &=&
    +\fracown{19}{1890}
    \,,
  \nonumber\\
  a_{30}^{7c} &=&
    -\fracown{4\pi^2}{3645}
    -\fracown{752008}{24111675}
    \,,
  \nonumber\\
  a_{31}^{7c} &=&
    -\fracown{5876}{382725}
    \,,
  \nonumber\\
  a_{32}^{7c} &=&
    -\fracown{4}{1215}
    \,,
  \nonumber\\
  a_{40}^{7c} &=&
    +\fracown{3075421}{1296672300}
    \,,
  \nonumber\\
  a_{41}^{7c} &=&
    +\fracown{74}{93555}
    \,,
\end{eqnarray}
\begin{eqnarray}
  b_{0}^{7c} &=&
    -\fracown{7048\zetathree}{243}
    -\fracown{69\Stwo}{14}
    +\fracown{103925\pi^2}{15309}
    -\fracown{6854384}{3444525}
    \,,
  \nonumber\\
  b_{1}^{7c} &=&
    -\fracown{448\zetathree}{27}
    +\fracown{653\Stwo}{35}
    -\fracown{29\pi^2}{8505}
    +\fracown{670214}{42525}
    \,,
  \nonumber\\
  b_{2}^{7c} &=&
    -\fracown{224\zetathree}{9}
    +\fracown{10109\Stwo}{210}
    +\fracown{\pi^2}{17010}
    +\fracown{16246}{945}
    \,,
  \nonumber\\
  b_{3}^{7c} &=&
    -\fracown{896\zetathree}{27}
    +\fracown{3676\Stwo}{45}
    +\fracown{4\pi^2}{3645}
    +\fracown{7423}{405}
    \,,
  \nonumber\\
  b_{4}^{7c} &=&
    -\fracown{1120\zetathree}{27}
    +\fracown{3166\Stwo}{27}
    +\fracown{92539}{4860}
    \,,
  \nonumber\\
  b_{5}^{7c} &=&
    -\fracown{448\zetathree}{9}
    +\fracown{62512\Stwo}{405}
    +\fracown{352919}{18225}
    \,,
  \nonumber\\
  b_{6}^{7c} &=&
    -\fracown{1568\zetathree}{27}
    +\fracown{5200\Stwo}{27}
    +\fracown{235619}{12150}
    \,,
  \nonumber\\
  b_{7}^{7c} &=&
    -\fracown{1792\zetathree}{27}
    +\fracown{1970884\Stwo}{8505}
    +\fracown{1468651}{76545}
    \,,
  \nonumber\\
  b_{8}^{7c} &=&
    -\fracown{224\zetathree}{3}
    +\fracown{13859809\Stwo}{51030}
    +\fracown{6903803}{367416}
    \,,
\end{eqnarray}
\begin{eqnarray}
  a_{00}^{8c} &=&
    -\fracown{1627\zetathree}{162}
    +\fracown{25583\pi^2}{5832}
    +\fracown{9148337}{104976}
    \,,
  \nonumber\\
  a_{01}^{8c} &=&
    +\fracown{1049}{486}
    \,,
  \nonumber\\
  a_{02}^{8c} &=&
    -\fracown{23}{81}
    \,,
  \nonumber\\
  a_{10}^{8c} &=&
    +\fracown{9707}{32400}
    \,,
  \nonumber\\
  a_{11}^{8c} &=&
    +\fracown{601}{2160}
    \,,
  \nonumber\\
  a_{20}^{8c} &=&
    +\fracown{29\pi^2}{7560}
    +\fracown{462793}{6945750}
    \,,
  \nonumber\\
  a_{21}^{8c} &=&
    +\fracown{2479}{66150}
    \,,
  \nonumber\\
  a_{22}^{8c} &=&
    +\fracown{29}{2520}
    \,,
  \nonumber\\
  a_{30}^{8c} &=&
    +\fracown{11\pi^2}{108864}
    -\fracown{1616438903}{144027072000}
    \,,
  \nonumber\\
  a_{31}^{8c} &=&
    -\fracown{496373}{114307200}
    \,,
  \nonumber\\
  a_{32}^{8c} &=&
    +\fracown{11}{36288}
    \,,
  \nonumber\\
  a_{40}^{8c} &=&
    -\fracown{13282901}{8644482000}
    \,,
  \nonumber\\
  a_{41}^{8c} &=&
    -\fracown{1879}{2494800}
    \,,
\end{eqnarray}
\begin{eqnarray}
  b_{0}^{8c} &=&
    -\fracown{2833\zetathree}{162}
    -\fracown{29871\Stwo}{2240}
    +\fracown{7169669\pi^2}{1632960}
    +\fracown{7342571207}{73483200}
    \,,
  \nonumber\\
  b_{1}^{8c} &=&
    -\fracown{131\zetathree}{9}
    +\fracown{597\Stwo}{2240}
    -\fracown{1447\pi^2}{181440}
    +\fracown{17694343}{907200}
    \,,
  \nonumber\\
  b_{2}^{8c} &=&
    -\fracown{65\zetathree}{3}
    +\fracown{50339\Stwo}{2240}
    +\fracown{751\pi^2}{181440}
    +\fracown{1260331}{60480}
    \,,
  \nonumber\\
  b_{3}^{8c} &=&
    -\fracown{259\zetathree}{9}
    +\fracown{64417\Stwo}{1344}
    -\fracown{11\pi^2}{108864}
    +\fracown{4083773}{181440}
    \,,
  \nonumber\\
  b_{4}^{8c} &=&
    -\fracown{323\zetathree}{9}
    +\fracown{10871\Stwo}{144}
    +\fracown{307651}{12960}
    \,,
  \nonumber\\
  b_{5}^{8c} &=&
    -43\zetathree
    +\fracown{45223\Stwo}{432}
    +\fracown{4783799}{194400}
    \,,
  \nonumber\\
  b_{6}^{8c} &=&
    -\fracown{451\zetathree}{9}
    +\fracown{2431\Stwo}{18}
    +\fracown{816431}{32400}
    \,,
  \nonumber\\
  b_{7}^{8c} &=&
    -\fracown{515\zetathree}{9}
    +\fracown{3773167\Stwo}{22680}
    +\fracown{10434863}{408240}
    \,,
  \nonumber\\
  b_{8}^{8c} &=&
    -\fracown{193\zetathree}{3}
    +\fracown{108012589\Stwo}{544320}
    +\fracown{252135383}{9797760}
    \,.
\end{eqnarray}
\mathindent1cm

\setlength {\baselineskip}{0.2in}
 

\begin{thebibliography}{99}
%
\bibitem{Chen:2001fj}
S.~Chen {\it et al.}  (CLEO Collaboration),
Phys.\ Rev.\ Lett.\  {\bf 87} (2001) 251807
[hep-ex/0108032].
%
\bibitem{Barate:1998vz}
R.~Barate {\it et al.}  (ALEPH Collaboration),
Phys.\ Lett.\ B {\bf 429} (1998) 169.
%
\bibitem{Abe:2001hk}
K.~Abe {\it et al.}  (BELLE Collaboration),
Phys.\ Lett.\ B {\bf 511} (2001) 151
[hep-ex/0103042].
%
\bibitem{Aubert:2002pd}
B.~Aubert {\it et al.}  (BABAR Collaboration),
[hep-ex/0207076].
%
\bibitem{Jessop.2002} C.~Jessop, SLAC report SLAC-PUB-9610, November 2002.
%
\bibitem{Gambino:2001ew}
P.~Gambino and M.~Misiak,
Nucl.\ Phys.\ B {\bf 611} (2001) 338
[hep-ph/0104034].
%
\bibitem{Buras:2002tp}
A.J.~Buras, A.~Czarnecki, M.~Misiak and J.~Urban,
Nucl.\ Phys.\ B  {\bf 631} (2002) 219
[hep-ph/0203135].
%
\bibitem{Greub:1996tg}
C.~Greub, T.~Hurth and D.~Wyler,
Phys.\ Rev.\ D {\bf 54} (1996) 3350
[hep-ph/9603404].
%
\bibitem{Bobeth:1999mk}
C.~Bobeth, M.~Misiak and J.~Urban,
Nucl.\ Phys.\ B {\bf 574} (2000) 291
[hep-ph/9910220].
%
\bibitem{Buras:2002er}
A.~J.~Buras and M.~Misiak,
Acta Phys. Pol. B 33 (2002) 2597
[hep-ph/0207131].
%
\bibitem{Bieri:2003ue}
K.~Bieri, C.~Greub and M.~Steinhauser,
Phys.\ Rev.\ D {\bf 67} (2003) 114019
[hep-ph/0302051].
%
\bibitem{GGH04} P.~Gambino, M.~Gorbahn and U.~Haisch, in preparation.
%
\bibitem{Steinhauser:2000ry}
M.~Steinhauser,
Comput.\ Phys.\ Commun.\  {\bf 134} (2001) 335
[hep-ph/0009029].
%
\bibitem{Chetyrkin:1996vx}
K.G.~Chetyrkin, M.~Misiak and M.~M\"unz,
Phys.\ Lett.\ B {\bf 400} (1997) 206,
Phys.\ Lett.\ B {\bf 425} (1998) 414 (E)
[hep-ph/9612313].
%
\bibitem{Gambino:2003zm}
P.~Gambino, M.~Gorbahn and U.~Haisch,
Nucl.\ Phys.\ B {\bf 673} (2003) 238
[hep-ph/0306079].
%
\bibitem{Grinstein:tj}
B.~Grinstein, R.~P.~Springer and M.~B.~Wise,
Nucl.\ Phys.\ B {\bf 339} (1990) 269.
%
\bibitem{Smi02}
V.~A. Smirnov,
{\it Applied Asymptotic Expansions in Momenta and Masses},
Springer-Verlag, Heidelberg, 2001.
%
\bibitem{Nogueira:1993ex}
P.~Nogueira,
J.~Comp.~Phys. {\bf 105} (1993) 279.
%
\bibitem{q2e}
T.~Seidensticker, unpublished.
%
\bibitem{exp}
T.~Seidensticker,
hep-ph/9905298;\\
R.~Harlander, T.~Seidensticker and M.~Steinhauser,
Phys.\ Lett.\ B {\bf 426} (1998) 125
[hep-ph/9712228].
%
\bibitem{Form}
J.A.M. Vermaseren, {\it Symbolic Manipulation with FORM},
  Computer Algebra Netherlands, Amsterdam, 1991.
%
\bibitem{Kublbeck:xc}
J.~K\"ublbeck, M.~B\"ohm and A.~Denner,
Comput.\ Phys.\ Commun.\  {\bf 60} (1990) 165;\\
T.~Hahn and C.~Schappacher,
Comput.\ Phys.\ Commun.\  {\bf 143} (2002) 54
[hep-ph/0105349].
%
\bibitem{Steinhauser:2002rq}
M.~Steinhauser,
Phys.\ Rept.\  {\bf 364} (2002) 247
[hep-ph/0201075].

\end{thebibliography}
\end{document}